\documentclass[apj]{emulateapj}
\slugcomment{{\sc Accepted to ApJS:} December 20, 2016}
\def\gtorder{\mathrel{\raise.3ex\hbox{$>$}\mkern-14mu
                \lower0.6ex\hbox{$\sim$}}}
\def\ltorder{\mathrel{\raise.3ex\hbox{$<$}\mkern-14mu
                \lower0.6ex\hbox{$\sim$}}}

\shorttitle{Cooling in Photoionized Plasma}
\shortauthors{O. Gnat}

\begin{document}
\title{Time-Dependent Cooling in Photoionized Plasma}
\vspace{1cm}
\author{Orly Gnat\altaffilmark{1}}
\altaffiltext{1}{Racah Institute of Physics, The Hebrew University, Jerusalem 91904, Israel}
\email{orlyg@phys.huji.ac.il}

\begin{abstract}
I explore the thermal evolution and ionization states in 
gas cooling from an initially hot state in the presence of external
photoionizing radiation. 
I compute the equilibrium and nonequilibrium cooling 
efficiencies, heating rates, and ion fractions for low-density gas cooling while exposed 
to the ionizing metagalactic background radiation at various redshifts
($z=0-3$), for a range of temperatures ($10^8$ - $10^4$~K), densities 
($10^{-7}-10^3$~cm$^{-3}$) and metallicities ($10^{-3}$ - $2$ times solar). 
The results indicate the existence of a threshold 
ionization parameter, above which the cooling efficiencies are very close 
to those in photoionization equilibrium (so that departures from equilibrium
may be neglected), and below which the cooling efficiencies resemble those 
in collisional time-dependent gas cooling with no external radiation (and 
are thus independent of density). 
\end{abstract}


\section{Introduction}
\label{introduction}

The radiative cooling efficiencies of low-density plasma are necessary quantities
in the study of diffuse gas in and around galaxies. They are important in setting 
the rates at which gas accretes onto, and is ejected from, forming galaxies, and are 
therefore essential for modeling various aspects of galaxy formation and evolution.

The time-dependent radiative cooling of highly ionized gas,
in the absence of any external sources of heating or photoionization,
is a well-studied problem dating back to 
Kafatos (1973),
that has been revisited by many other authors.
Here I refer to such conditions as ``collisional'', as opposed to 
``photoionized'' gas. In photoionized gas non-collisional (i.e. radiative) 
processes  become significant.
Gnat \& Sternberg (2007, hereafter GS07) studied the time-dependent ionization
and cooling in collisional plasma. They calculated the equilibrium and nonequilibrium 
ionization states and radiative cooling efficiencies of low density gas cooling 
from an initial hot temperature. 
With no radiation, the collisional ionization {\it equilibrium} (CIE) ion 
fractions and radiative cooling efficiencies are functions of the temperature 
only.

In GS07 the main emphasis has been on the departures from equilibrium
that occur when the cooling time becomes short compared with the
electron-ion recombination time in collisional plasma. 
Below $\sim5\times10^6$~K,
recombination becomes slow compared with cooling, and so
the plasma tends so remain more highly ionized than expected for CIE.
More highly ionized species generally have more energetic resonance 
line transitions, and so thermal electrons are less efficient in
exciting over-ionized plasma, causing a suppression of the cooling 
efficiency at any temperature. Because metal-lines dominate the cooling
over a wide range of temperatures, departures from equilibrium are
more prominent for high-metallicity gas.

In astrophysical environments, low-density gas rarely exists in the absence of 
significant sources of ionizing radiation, either stellar or extragalactic.
This radiation modifies the ionization states, thus directly 
affecting the observational signatures. 
For example, photoionizing radiation is key to understanding 
the absorption line signatures from circum-galactic
gas (Churchill et al.~2015; Crighton et al.~2015);
the origin of cold gas in galactic winds (Thompson et al.~2015);
and the physical properties of gas around star-forming galaxies (Turner et al.~2015);

Crucially, when photoionization modifies the ionization states, it affects 
the cooling efficiencies which in turn determine the rates and epochs at 
which galaxies accrete their gas (this idea dates back to white \& Rees~1978).
At high redshifts, where the IGM is composed of primordial gas, photoionizing 
radiation has been shown to suppresses cooling significantly, thus quenching the 
formation rate of low-mass galaxies (e.g. Efstathiou~1992).
More recent studies have shown that higher-mass galaxies also have their
gas accretion- and star formation-rates regulated by the photoionizing
radiation (e.g., Navvaro \& Steinmetz~1997; 
Benson et al.~2002a; 2002b; 2003; 
Cantalupo~2010. 
See, however, Weinberg et al.~1997).
The photoionized cooling efficiencies are therefore key to understanding 
both galaxy- and star-formation.

The ionization states in photoionization equilibrium have been studied
extensively (E.g. recently by Churchill et al.~2014).
The cooling and heating functions in photoionization equilibrium have been 
recently studied by Gnedin \& Hollon (2012; for stellar and AGN spectral energy 
distributions) and Wiersma et al. (2009; for the CMB and metagalactic background).
The nonequilibrium radiative cooling of hot ($>10^4$~K) {\it photoionized} gas, 
and the associated time-dependent ionization of the metal ions have been recently
considered by Vasiliev (2011) and by Oppenheimer \& Schaye (2013).
Oppenheimer \& Schaye (2013) presented results for gas densities
between  $10^{-5}$ and $10^{-2}$~cm$^{-3}$, and for gas metallicities
between $0.1$ and $2$ times solar.
Departures from equilibrium cooling has been shown to affect the temperature
of the IGM (Punchwein et al. 2015); the UV absorption signatures in the proximity
of AGN (Oppenheimer \& Schaye~2013); and the composition of galactic outflows
(Richings \& Schaye~2015); 

In this paper, I reexamine the fundamental problem of nonequilibrium 
ionization of a time-dependent radiatively cooling gas in the presence 
of external photoionization. 
I focus on the impact that the photoionizing background has on the 
thermal evolution and ionization states. I consider the metagalactic 
background radiation (Haardt \& Madau~2012, hereafter HM12) 
for redshifts between $0$ and $3$,
gas densities between $10^{-7}$ and $10^3$~cm$^{-3}$, and metallicities
between $10^{-3}$ and $2$ times solar.
When photoionization and heating by the metagalactic background affect the 
physical properties of the cooling gas, the conditions become functions 
of the temperature, metallicity {\it and density} (or, equivalently,
ionization parameter) of the gas.  

The computations presented 
here indicate the existence of a threshold ionization parameter, above which the 
nonequilibrium cooling efficiencies are very close to those in photoionization 
equilibrium (so that departures from equilibrium may be neglected), and below which 
the cooling efficiencies resemble those in collisional time-dependent gas cooling 
with no external radiation (and are thus independent of density). 

The outline of this paper is as follows. In \S2 (and Appendix~\ref{numericaldetails})
I summarize the numerical method and physical ingredients included in this computation.
In \S3 I describe how the photoionizing radiation affects the thermal evolution of
radiatively cooling gas. I compare the equilibrium and nonequilibrium cooling 
efficiencies, and identify the threshold ionization parameter that separates
the regime of photoionization equilibrium from the regime of time-dependent
collisional cooling. \S4 presents the ionization states as functions of gas 
temperature and density, for both photoionization equilibrium and nonequilibrium 
cooling. The full set of results, 
for metallicities $10^{-3}$ - $2$ times solar, and for densities between
$10^{-7}$ and $10^3$~cm$^{-3}$, are available as online data files. 
In \S5 I discuss the evolution of the density ratios 
C~{\footnotesize IV}$/$O~{\footnotesize VI} versus 
N~{\footnotesize V}$/$O~{\footnotesize VI} in photoionized cooling gas.
This density ratio can be used as a diagnostic tool for photoionized plasma.
Additional diagnostic diagrams can be constructed using the online data presented
in this paper. I summarize in \S6.

\section{Numerical Method and Processes}
\label{method}

The goal of this paper is to study how an external radiation field affects
the observational signatures in radiatively cooling plasma.
I consider gas which is initially heated to a temperature $\gtrsim5\times10^6$~K, 
and then cools radiatively. As in GS07, at the initial hot state, cooling
is slower than recombination, and so the gas can reach ionization equilibrium
before significant cooling takes place.
As opposed to the situation discussed in GS07, here the gas is exposed to a 
continuous and constant source of heating and photoionization as it cools. 
The initial ionization equilibrium state is that of {\it photo}ionization 
equilibrium (as opposed to CIE in GS07). 
In all cases considered here, the heating rate at
the initial photoionization equilibrium state is lower than the cooling rate,
and therefore the gas is not in {\it thermal} equilibrium, and will cool radiatively
until the condition of thermal equilibrium is satisfied.
I compute the coupled time-dependent evolution for clouds cooling at constant density
(see \S3.1 for a discussion of isobaric evolution), for 
a wide range of metallicities, taking into account departures from photoionization
equilibrium.

In my computations I follow the numerical scheme outlined in 
GS07 and improved in Gnat \& Sternberg (2009). The details of this numerical
method and computational ingredients are summarized in 
Appendix~\ref{numericaldetails}.
I consider all ionization stages of the 
elements H, He, C, N, O, Ne, Mg, Si, S, and Fe. I include photoionization,
collisional ionization by thermal electrons, radiative recombination, 
dielectronic recombination,  neutralization and ionization by charge transfer
reactions with hydrogen and helium atoms and ions, and multi electron
Auger ionization processes.
The photoionization rates are due to the externally
incident radiation.

The ionization equations are coupled to an energy
equation for the time-dependent heating and 
cooling, and resulting temperature variation.
I follow the electron cooling
efficiency, $\Lambda(T,x_i,Z)$ 
(erg~s$^{-1}$~cm$^3$), and the heating rate $\Upsilon(J_{\nu},x_i,Z)$
(erg~s$^{-1}$),
which depend on the gas temperature, 
the photoionizing background, the
ionization state, and the metallicity $Z$.
As in GS07, I adopt the elemental abundances reported by
Asplund et al. (2005) for the photosphere of the Sun, and the 
enhanced
Ne abundance
recommended by Drake \& Testa (2005; see Table~1 in GS07). 
In all computations I assume a primordial
helium abundance $A_{\rm He}=1/12$ (Ballantyne et al.~2000),
independent of $Z$. 

The electron cooling efficiency
includes the removal of electron {\it kinetic} energy\footnote{I 
do not include the ionization potential
energies as part of the internal energy 
but instead follow the loss and gain
of the electron kinetic energy only.}
via recombination with ions, collisional 
ionization, collisional excitation followed by
prompt line emission, thermal bremsstrahlung (GS07),
and Compton cooling off the metagalactic background
radiation.
The compton cooling power is given by 
\begin{equation}
P_{\rm compton} = \left(\frac{4kT}{m_e c}\right)\sigma_T n_e \Phi_{\rm ph},
\label{compton}
\end{equation}
where,
\begin{equation}
\Phi_{\rm ph} = \frac{4\pi}{c}\int_{0}^{\infty}{J_\nu d\nu}.
\end{equation}
Most of the cooling processes (line-emission, recombinations,
ionization, thermal bremsstrahlung) are two-body
processes, and are therefore proportional to the density squared
for the low densities considered here ($10^{-7}-10^3$~cm$^{-3}$).
However, Compton cooling is proportional to the electron density times the 
photon density, and therefore dominates at low densities, where all other 
processes are suppressed.

For the heating
rate, $\Upsilon(x_i,Z,J_{\nu})$ (erg~s$^{-1}$)
I include compton heating by high energy photons and 
photo-ionization by the background radiation,
$J_\nu$. Each photoionizing photon adds an energy
$h\nu_\gamma - h\nu_{\rm IP}$ to the kinetic energy of
the electron gas, where $\nu_\gamma$ is the absorbed photon
frequency, and $\nu_{\rm IP}$ is the ionization threshold 
of the ionized ion.

The cooling efficiencies and heating rates were
computed by passing the nonequilibrium ion fractions $x_i(T)$ to the 
Cloudy cooling and heating functions. 
The results presented here have been obtained 
using Cloudy version 13.00.  
The net local cooling rate per volume is given
by $n_e n_H \Lambda_{\rm e.H} - n_H\Upsilon$.

For an ideal gas, the pressure $P=nk_{\rm B}T$,
and the thermal energy density $u=3/2nk_{\rm B}T$.
If $dQ$ is the amount of heat lost (or gained) by the 
thermal electron gas, then for isochoric cooling (for which
$dV=0$), 
\begin{equation}
dQ = dU = \frac{3}{2} (Nk_BdT+k_BTdN)
\end{equation}

This leads to the relation,
\begin{equation}
\frac{3}{2} \frac{dP}{dt} = -n_e n_{\rm H} \Lambda(T,x_i,Z) + n_H \Upsilon(J_{\nu},x_i,Z)
\label{energy}
\end{equation}
(GS07, Kafatos 1973). 

In the absence of external radiation (i.e. when $\Gamma_i = 0$ and
$\Upsilon=0$) the 
evolution of the ion fractions as functions of temperature, 
$\frac{dx_i}{dT} = (\frac{dx_i}{dt})/(\frac{dT}{dt})$, is 
independent of the gas density or pressure.
When external radiation is present, the ion fractions
at a given temperature and ionizing background
are functions of the gas density or,
equivalently, of the ionization parameter,
\begin{equation}
U = \frac{4\pi}{n_e c} \int_{\nu_0}^{\infty}\frac{J_\nu}{h\nu}d\nu
\label{IP}
\end{equation}
which measures the ratio of the ionizing photon density
to the electron density.

The ionic rate equations 
and energy-balance equation $($\ref{energy}$)$ constitute a set 
of $103$ coupled ordinary differential equations (ODEs).
I use the numerical schemed outlined in GS07 for isochorically
cooling gas to solve for the non-equilibrium ion-fractions and
temperature. Here the local errors for hydrogen, 
helium, and metal ions were set to be smaller than $10^{-9}$, $10^{-8}$, 
and $10^{-7}$, respectively. The high accuracy is crucial, because
heating is sometimes dominated by trace species. 
The integration has been carried out 
to a minimum temperature $T_{\rm low}$, set by the condition of thermal
equilibrium with the background radiation field.
If the equilibrium temperature is lower than $10^4$~K,
I set $T_{\rm low} = 10^4$~K, because molecular chemistry and dust cooling are 
not included in this work, and the results for temperatures below
$10^4$~K may therefore be unreliable.

\section{Cooling Efficiencies}
\label{cooling}

In this section I describe the thermal evolution of the cooling gas, 
and examine its dependence on the controlling parameters: the 
ionizing radiation, gas density, and metallicity.

\begin{figure}
\epsscale{1.2}
\plotone{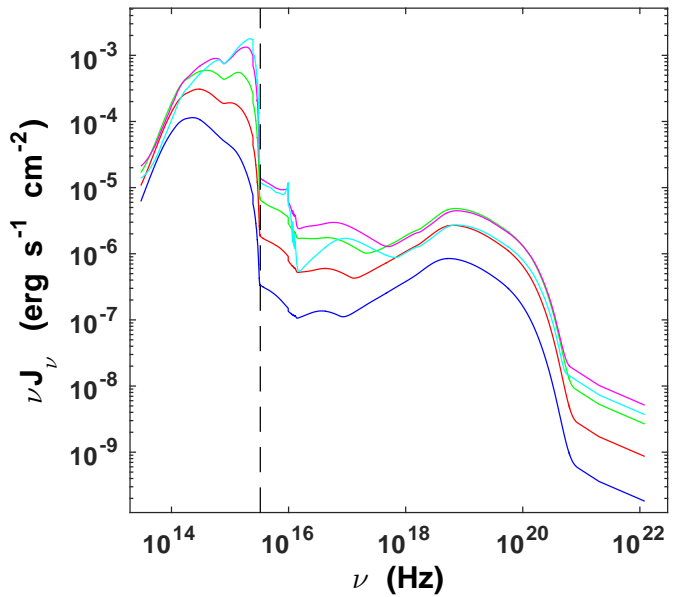}
\caption{Mean intensity versus frequency in the
Haardt \& Madau (2012) background radiation
at $z=0$, $0.5$, $1$, $2$, and $3$. The mean intensity 
of ionizing photons rises
from $z=0$ to $z=2$, and then decreases for higher redshifts.
The dashed curve shows the Lyman limit.}
\label{HM}
\end{figure}

For the photoionizing radiation field, I consider the metagalactic 
background radiation at redshifts, $z=0,~0.5,~1,~2,$ and $3$ 
(HM12), between $0.009$ and $\sim6\times10^5$~Ryd ($\sim0.1$~eV$-8,330$~keV). 
I show the full HM12 spectral energy distributions 
in Figure~\ref{HM}.
The mean intensity of the metagalactic background increases between
redshift $0$ and $2$. It reaches a maximum at $z\sim2$ and then
decreases again at higher redshifts. The ionizing photon densities 
($4\pi/c \int{J_\nu/h\nu~d\nu}$) are listed in Table~\ref{photons}.

\begin{deluxetable}{lc}
\tablewidth{0pt}
\tablecaption{Ionizing Photon Density in HM12 SEDs}
\tablehead{
\colhead{Redshift} & 
\colhead{$n_{\gamma}=U~n_H$}}
\startdata
$0$   & $4.2\times10^{-6}$ \\
$0.5$ & $2.2\times10^{-5}$ \\
$1$   & $7.8\times10^{-5}$ \\
$2$   & $1.6\times10^{-4}$ \\
$3$   & $1.4\times10^{-4}$ \\
\enddata
\label{photons}
\end{deluxetable}

I consider gas densities between $10^{-7}$~cm$^{-3}$
($\sim$the cosmic mean) and $10^3$~cm$^{-3}$,
and explore five different values of gas metallicity
from $10^{-3}$ to $2$ times the metal abundance of the sun.
For each redshift-density-metallicity combination, I have carried
out computations of the ion fractions, cooling efficiencies and heating
rates as functions of gas temperature.
First, I assume photoionization equilibrium (hereafter PIE, see Table~\ref{glossary})
imposed at all temperatures
to calculate the PIE ion fractions and cooling
efficiencies.
Then I consider the non-equilibrium evolution of photoionized cooling gas,
to compute the time-dependent ionization and cooling.

\begin{deluxetable*}{lcc}
\tablewidth{0pt}
\tablecaption{Acronym Glossary}
\tablehead{
\colhead{Processes} & 
\colhead{Equilibrium} &
\colhead{Time-Dependent} }
\startdata
Collisional Only    & CIE & TDC \\
& {\scriptsize Collisional Ionization Equilibrium} & {\scriptsize Time-Dependent Collisional} \\
\\
Including Radiation & PIE & TDP \\
& {\scriptsize Photo-Ionization Equilibrium} & {\scriptsize Time-Dependent Photoionized} \\
\enddata
\label{glossary}
\end{deluxetable*}

\begin{figure*}
\epsscale{1.2}
\plotone{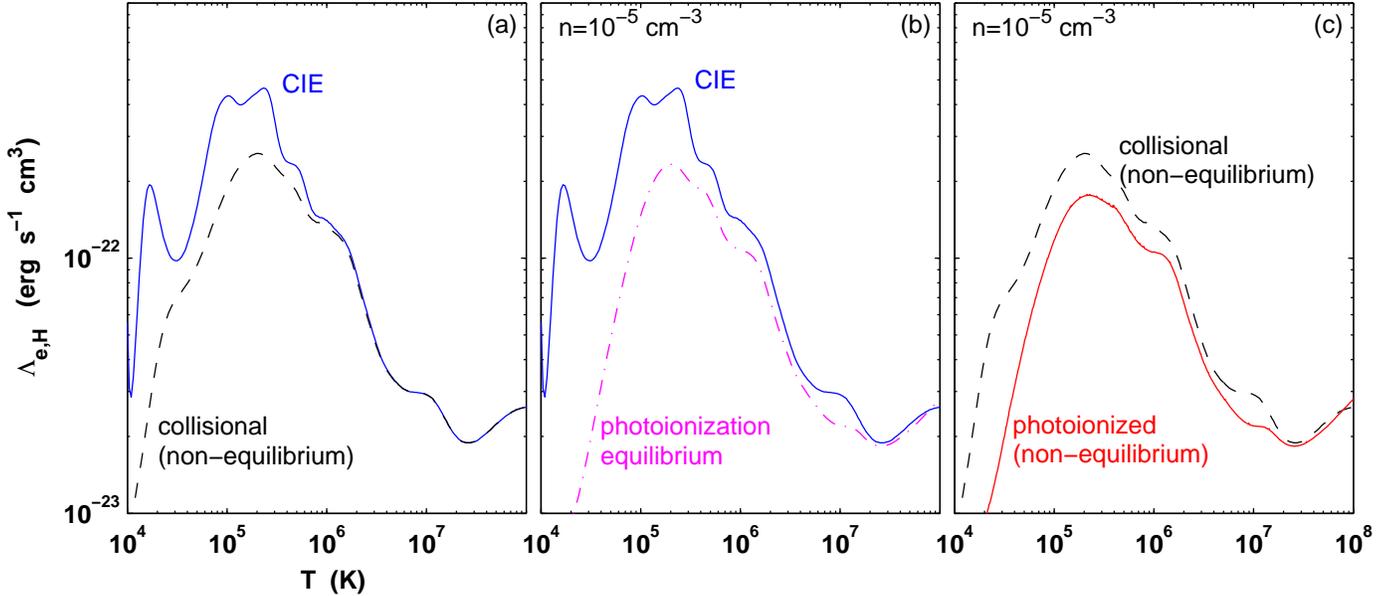}
\caption{Cooling efficiencies (erg s$^{-1}$ cm$^3$) versus temperature.
(a) CIE cooling (solid blue) versus time-dependent collisional cooling (dashed black).
Data from GS07. 
(b) CIE cooling (solid blue) versus PIE 
cooling
(dash-dotted magenta curve) assuming a density of $10^{-5}$~cm$^{-3}$,
and the $z=0$ HM12 background;
(c) 
TDC cooling (no radiation, dashed black)
versus time-dependent cooling in photoionized gas with a density
of $10^{-5}$~cm$^{-3}$ (solid red).}
\label{z0Z0}
\end{figure*}

Here I focus on the total cooling efficiencies, 
$\Lambda_{H,e}(z, Z, T)$~(ergs~cm$^3$~s$^{-1}$), in the photoionized
cooling gas. Figure~\ref{z0Z0} shows the cooling efficiencies as a function
of gas temperature for solar metallicity gas. Panel (a) shows the cooling 
for collisional plasma, i.e., with no external radiation (GS07). 
The blue solid curve is for CIE. This familiar curve shows the
hydrogen Ly$\alpha$ peak at $\sim10^4$~K, followed by dominant 
contributions from carbon, oxygen, neon and iron resonance-lines, and
thermal bremsstrahlung (from low to high temperature).
The black dashed curve is for time-dependent collisional cooling (hereafter TDC,
see Table~\ref{glossary}). 
The narrow contributions dominated by individual species that appear 
for CIE are smeared out in the 
TDC curves. This is due to the broader ion distribution
that occur as overionized species persist down to low temperatures.
The nonequilibrium collisional cooling efficiencies are suppressed, by factors of 
$2 - 4$, compared to CIE cooling. This is because the gas remains "overionized" as 
it cools, and consequently tends to have more energetic resonance line transitions
(McCray 1987), which are less accessible to the ``cooler'' thermal electrons.

Panel (b) compares the CIE cooling curve (again, solid blue) with the cooling 
efficiencies assuming PIE in a $10^{-5}$~cm$^{-3}$ gas
in the presence of the $z=0$ HM12 metagalactic radiation field (dash-dotted
magenta curve).
Photoionized gas is more highly ionized than gas in collisional equilibrium
because of the additional ionization term due to radiation. Just as in the case
of departures from equilibrium, the more highly ionized species are less efficiently
excited by the thermal electrons, and cooling is suppressed.

Finally, panel (c) compares the {\it non-equilibrium} cooling efficiencies in the 
collisional case (TDC, dashed black) and in the photoionized case (solid red) 
assuming $10^{-5}$~cm$^{-3}$ gas in the presence of the $z=0$ HM12 metagalactic 
radiation field. For this gas density, even though the time-dependent collisional
gas is overionized compared with CIE, the addition of photoionization by the
metagalactic background increases the ionization states of the dominant coolants 
above and beyond those in the non-equilibrium collisional gas.

\begin{figure}
\epsscale{1.2}
\plotone{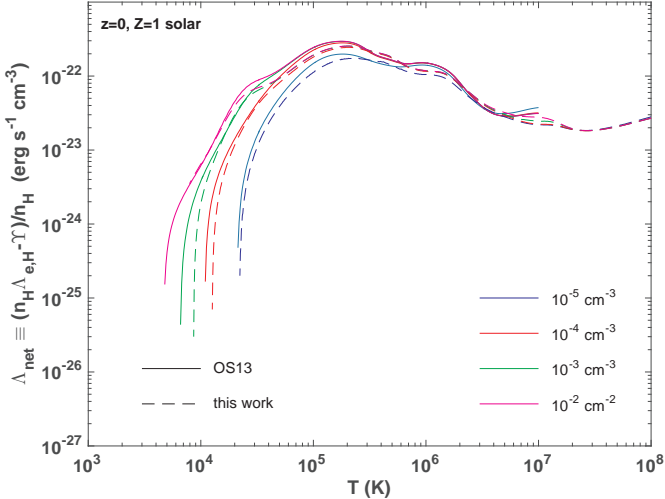}
\caption{Net cooling efficiencies ($\equiv(n_{\rm H}\Lambda_{\rm e,H}-\Upsilon)/n_{\rm H}$~erg s$^{-1}$ cm$^3$) 
versus temperature in this work (dashed curves) versus OS13 for solar 
metallicity gas exposed to the $z=$ radiation field. The different colors correspond 
to different gas densities.}
\label{OScomp}
\end{figure}

Figure~\ref{OScomp} compares the TDP cooling efficiencies computed here 
to those of 
Oppenheimer \& Schaye (2013, hereafter OS13). For a direct comparison,
the figure shows the {\it net} cooling efficiencies, defined here as
$\Lambda_{\rm net}\equiv(n_{\rm H}\Lambda_{\rm e,H}-\Upsilon)/n_{\rm H}$.
OS13 provide online results for densities in the range 
$n_{\rm H}=10^{-5}$~cm$^{-3}$
and $10^{-2}$~cm$^{-3}$, which are displayed in Figure~\ref{OScomp}. 
Note, that these densities are all ``high densities'' (i.e. above the 
threshold density) in the context of the discussion that follows. 
I display results for solar metallicity gas exposed
to the $z=0$ HM12 radiation field. Figure~\ref{OScomp} shows a very nice
agreement between the results of this work (dashed curves) and OS13 (solid
curves). The small differences 
seen between $10^5$~K and $\sim2\times10^6$~K have already been described
in OS13, and are mainly due to differences in the solar composition 
assumed (see figure~5 and associated discussion in OS13). 
The low temperatures differences are all basically at thermal equilibrium, 
where cooling and heating become comparable,
and the differences between them are minute. These results are also
very sensitive to the high-energy cutoff of the input SED (see discussion
in Appendix~\ref{xrayapp}). Overall, the agreement between the 
computed TDP cooling efficiencies presented in Figure~\ref{OScomp}, is better 
than the one presented in OS13 for the TDC rates, mainly due to the use of 
the more up-to-date 
atomic data included in the newer version of Cloudy (ver. 13.00). 

\begin{deluxetable}{lll}
\tablewidth{0pt}
\tablecaption{Cooling Data Guide\label{Tguide}}
\tablehead{
\colhead{Density (cm$^{-3}$)}&
\colhead{redshift}&
\colhead{Lettered Index}}
\startdata
$10^{-7}$ &  $3$   & A \\
$10^{-7}$ &  $2$   & B \\
$10^{-7}$ &  $1$   & C \\
$10^{-7}$ &  $0.5$ & D \\
$10^{-7}$ &  $0$   & E \\
$10^{-6}$ &  $3$   & F \\
$10^{-6}$ &  $2$   & G \\
$10^{-6}$ &  $1$   & H \\
$10^{-6}$ &  $0.5$ & I \\
$10^{-6}$ &  $0$   & J \\
$10^{-5}$ &  $3$   & K \\
$10^{-5}$ &  $2$   & L \\
$10^{-5}$ &  $1$   & M \\
$10^{-5}$ &  $0.5$ & N \\
$10^{-5}$ &  $0$   & O \\
$10^{-4}$ &  $3$   & P \\
$10^{-4}$ &  $2$   & Q \\
$10^{-4}$ &  $1$   & R \\
$10^{-4}$ &  $0.5$ & S \\
$10^{-4}$ &  $0$   & T \\
$10^{-3}$ &  $3$   & U \\
$10^{-3}$ &  $2$   & V \\
$10^{-3}$ &  $1$   & W \\
$10^{-3}$ &  $0.5$ & X \\
$10^{-3}$ &  $0$   & Y \\
$10^{-2}$ &  $3$   & Z \\
$10^{-2}$ &  $2$   & AA \\
$10^{-2}$ &  $1$   & AB \\
$10^{-2}$ &  $0.5$ & AC \\
$10^{-2}$ &  $0$   & AD \\
$10^{-1}$ &  $3$   & AE \\
$10^{-1}$ &  $2$   & AF \\
$10^{-1}$ &  $1$   & AG \\
$10^{-1}$ &  $0.5$ & AH \\
$10^{-1}$ &  $0$   & AI \\
$1      $ &  $3$   & AJ \\
$1      $ &  $2$   & AK \\
$1      $ &  $1$   & AL \\
$1      $ &  $0.5$ & AM \\
$1      $ &  $0$   & AN \\
$10^{1}$  &  $3$   & AO \\
$10^{1}$  &  $2$   & AP \\
$10^{1}$  &  $1$   & AQ \\
$10^{1}$  &  $0.5$ & AR \\
$10^{1}$  &  $0$   & AS \\
$10^{2}$  &  $3$   & AT \\
$10^{2}$  &  $2$   & AU \\
$10^{2}$  &  $1$   & AV \\
$10^{2}$  &  $0.5$ & AW \\
$10^{2}$  &  $0$   & AX \\
$10^{3}$  &  $3$   & AY \\
$10^{3}$  &  $2$   & AZ \\
$10^{3}$  &  $1$   & BA \\
$10^{3}$  &  $0.5$ & BB \\
$10^{3}$  &  $0$   & BC \\
\enddata
\tablecomments{A density-redshift index to lettered parts A$-$BC used in
Tables~\ref{cool-eq}-\ref{cool-neq}, and \ref{heat-eq}-\ref{heat-neq}.}
\end{deluxetable}

\begin{deluxetable*}{lccccc}
\tablewidth{0pt}
\tablecaption{Equilibrium Cooling Efficiencies (erg s$^{-1}$~cm$^{3}$)\label{cool_eq}}
\tablehead{
\colhead{Temperature (K)}&
\colhead{$\Lambda_{\rm e,H}(Z=10^{-3})$ }&
\colhead{$\Lambda_{\rm e,H}(Z=10^{-2})$ }&
\colhead{$\Lambda_{\rm e,H}(Z=10^{-1})$ }&
\colhead{$\Lambda_{\rm e,H}(Z=1)$ }&
\colhead{$\Lambda_{\rm e,H}(Z=2)$ }}
\startdata
$1.0000\times10^{8}$ & $1.24\times10^{21}$ & $1.24\times10^{21}$ & $1.24\times10^{21}$ & $1.24\times10^{21}$ & $1.23\times10^{21}$ \\
$9.5012\times10^{7}$ & $1.18\times10^{21}$ & $1.18\times10^{21}$ & $1.18\times10^{21}$ & $1.18\times10^{21}$ & $1.17\times10^{21}$ \\
$9.0273\times10^{7}$ & $1.12\times10^{21}$ & $1.12\times10^{21}$ & $1.13\times10^{21}$ & $1.12\times10^{21}$ & $1.11\times10^{21}$ \\
$8.5770\times10^{7}$ & $1.07\times10^{21}$ & $1.07\times10^{21}$ & $1.08\times10^{21}$ & $1.07\times10^{21}$ & $1.06\times10^{21}$ \\
$\vdots$  &  $\vdots$ &  $\vdots$ &  $\vdots$ &  $\vdots$ &  $\vdots$ \\
\enddata
\tablecomments{The complete version of this table is in 
the electronic edition of the Journal. The printed edition 
contains only a sample. 
The full table is divided into lettered parts A$-$BC, and
lists the equilibrium cooling efficiencies
for for gas densities between $10^{-7}$ and $10^3$~cm$^{-3}$, for 
the metagalactic backgrounds at redshifts $0-3$, and for
$Z=10^{-3}$, $10^{-2}$, $10^{-1}$, $1$, and $2$ times solar metallicity
gas (for a guide, see Table~\ref{Tguide}).}
\label{cool-eq}
\end{deluxetable*}

\begin{deluxetable*}{lccccc}
\tablewidth{0pt}
\tablecaption{Non-equilibrium Cooling Efficiencies (erg s$^{-1}$~cm$^{3}$)\label{cool_neq}}
\tablehead{
\colhead{Temperature (K)}&
\colhead{$\Lambda_{\rm e,H}(Z=10^{-3})$ }&
\colhead{$\Lambda_{\rm e,H}(Z=10^{-2})$ }&
\colhead{$\Lambda_{\rm e,H}(Z=10^{-1})$ }&
\colhead{$\Lambda_{\rm e,H}(Z=1)$ }&
\colhead{$\Lambda_{\rm e,H}(Z=2)$ }}
\startdata
$1.0000\times10^{+8}$ & $1.24\times10^{-21}$ & $1.24\times10^{-21}$ & $1.24\times10^{-21}$ & $1.24\times10^{-21}$ & $1.23\times10^{-21}$\\
$9.9000\times10^{+7}$ & $1.23\times10^{-21}$ & $1.23\times10^{-21}$ & $1.23\times10^{-21}$ & $1.23\times10^{-21}$ & $1.22\times10^{-21}$\\
$9.8010\times10^{+7}$ & $1.22\times10^{-21}$ & $1.22\times10^{-21}$ & $1.22\times10^{-21}$ & $1.22\times10^{-21}$ & $1.20\times10^{-21}$\\
$9.7030\times10^{+7}$ & $1.20\times10^{-21}$ & $1.20\times10^{-21}$ & $1.20\times10^{-21}$ & $1.21\times10^{-21}$ & $1.19\times10^{-21}$\\
$\vdots$  &  $\vdots$ &  $\vdots$ &  $\vdots$ &  $\vdots$ &  $\vdots$ \\
\enddata
\tablecomments{The complete version of this table is in 
the electronic edition of the Journal. The printed edition 
contains only a sample. 
The full table is divided into lettered parts A$-$BC, and
lists the non-equilibrium cooling efficiencies
for for gas densities between $10^{-7}$ and $10^3$~cm$^{-3}$, for 
the metagalactic backgrounds at redshifts $0-3$, and for
$Z=10^{-3}$, $10^{-2}$, $10^{-1}$, $1$, and $2$ times solar metallicity
gas (for a guide, see Table~\ref{Tguide}).}
\label{cool-neq}
\end{deluxetable*}

\begin{figure*}
\vspace{0.1cm}
\epsscale{1.2}
\plotone{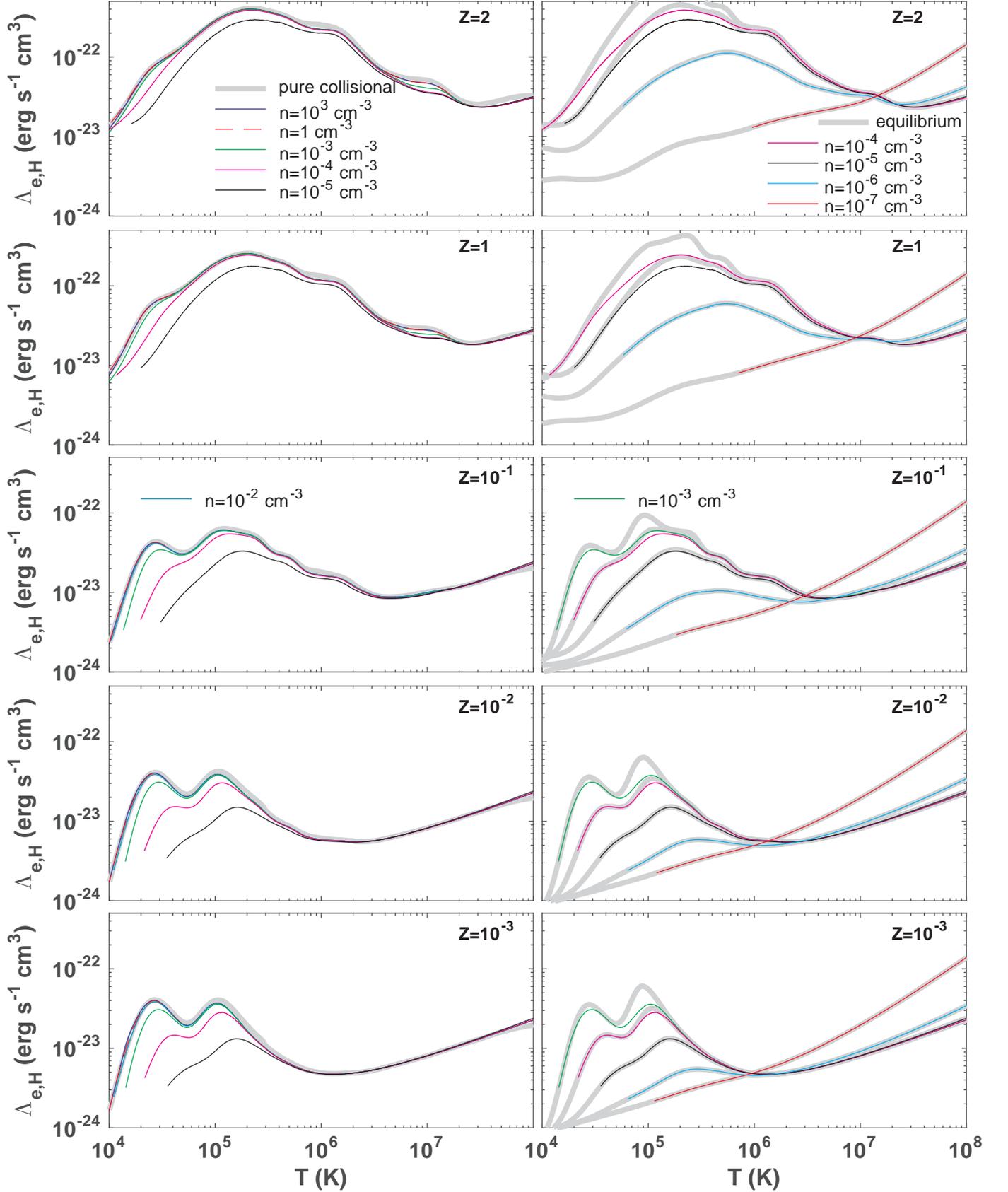}
\caption{\scriptsize{Time-dependent cooling efficiencies as functions of temperature 
in gas photoionized by the $z=0$
metagalactic radiation field, for gas densities between $10^3$ and $10^{-7}$~cm$^{-3}$
(shown by different colors), and gas metallicities between $2$ and $10^{-3}$ times
solar (top to bottom panels). The colored curves end at the thermal equilibrium
temperatures.
For higher density gas, $\gtrsim10^{-4}$~cm$^{-3}$ (shown in the left panels), the
cooling efficiencies are similar to those in collisional gas with no external radiation.
The time-dependent collisional cooling efficiency is independent of density, and is 
shown by the thick gray curve in the background.
For lower densities, $\lesssim10^{-4}$~cm$^{-3}$ (shown in the right panels), the
cooling efficiencies are similar to those of gas in photoionization {\it equilibrium},
shown by the series of gray curves behind the colored curves.}}
\label{fz0}
\end{figure*}

\begin{figure}
\epsscale{1.2}
\plotone{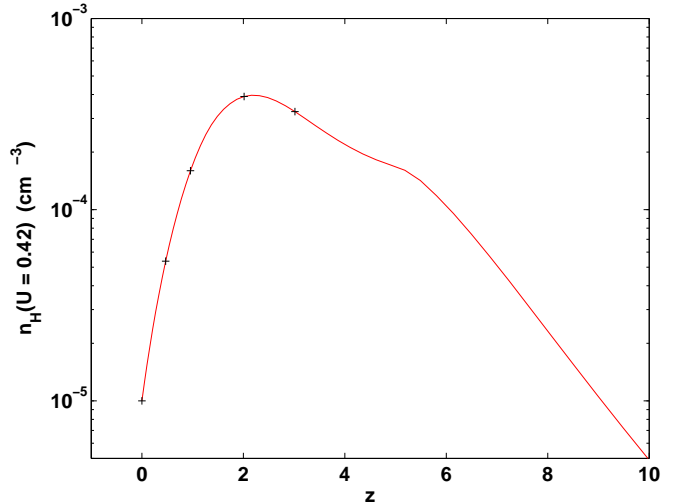}
\caption{Density for which $U=0.42$ as a function of redshift for the metagalactic
radiation field (Haardt \& Madau~2012). This is the density which marks the transition 
from {\it collisional} time-dependent cooling efficiencies
to photoionization {\it equilibrium} cooling efficiencies in solar metallicity gas ($Z\sim1$).
In primordial gas, this transition occurs at densities which are a factor $\sim100$ higher.}
\label{ncrit}
\end{figure}

\begin{figure*}
\vspace{0.3cm}
\epsscale{1.2}
\plotone{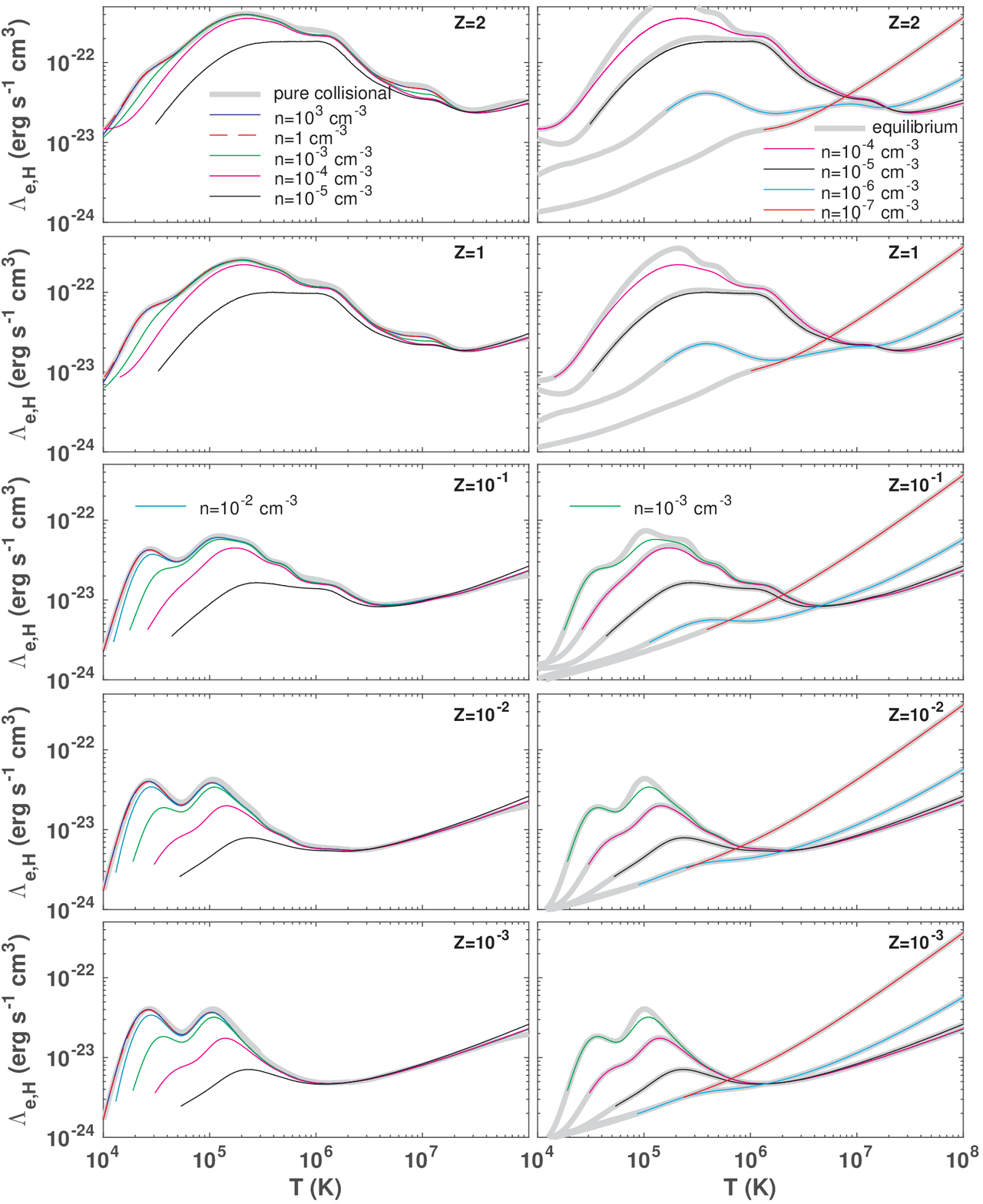}
\caption{Same as Figure~\ref{fz0}, but for the metagalactic radiation field at $z=0.5$.}
\label{fz05}
\end{figure*}

\begin{figure*}
\vspace{0.3cm}
\epsscale{1.2}
\plotone{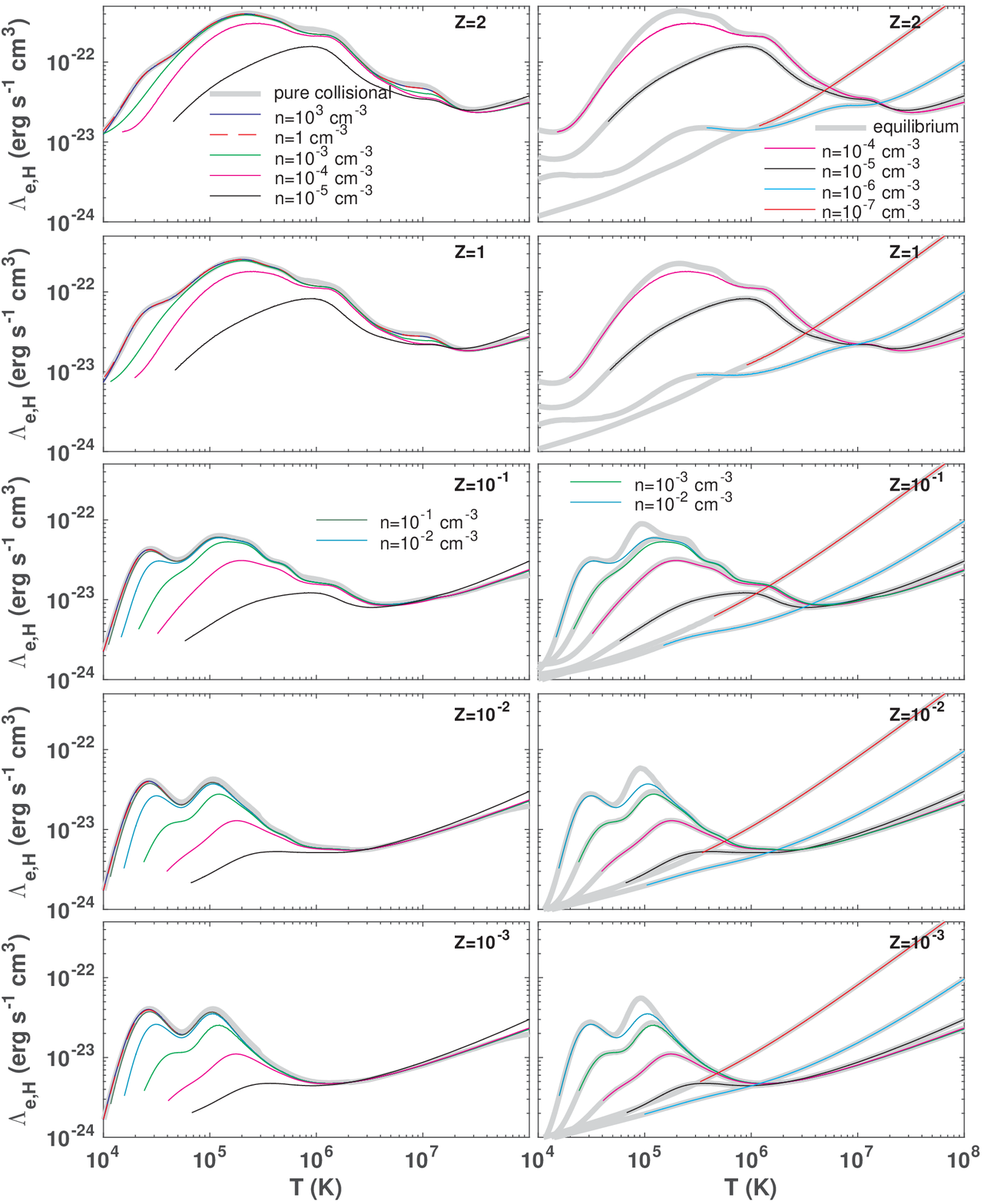}
\caption{Same as Figure~\ref{fz0}, but for the metagalactic radiation field at $z=1$.}
\label{fz1}
\end{figure*}

\begin{figure*}
\vspace{0.3cm}
\epsscale{1.2}
\plotone{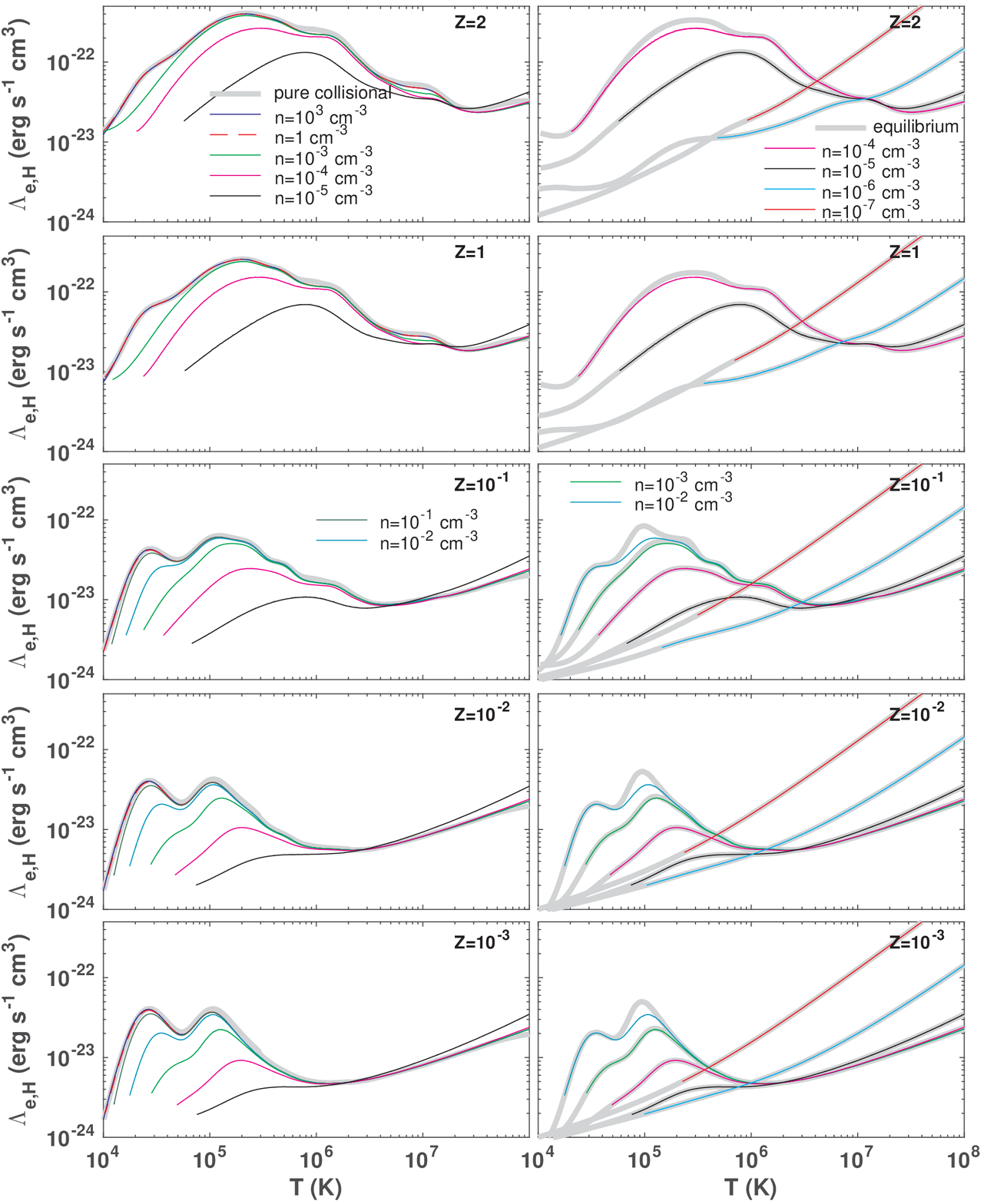}
\caption{Same as Figure~\ref{fz0}, but for the metagalactic radiation field at $z=2$.}
\label{fz2}
\end{figure*}

\begin{figure*}
\vspace{0.3cm}
\epsscale{1.2}
\plotone{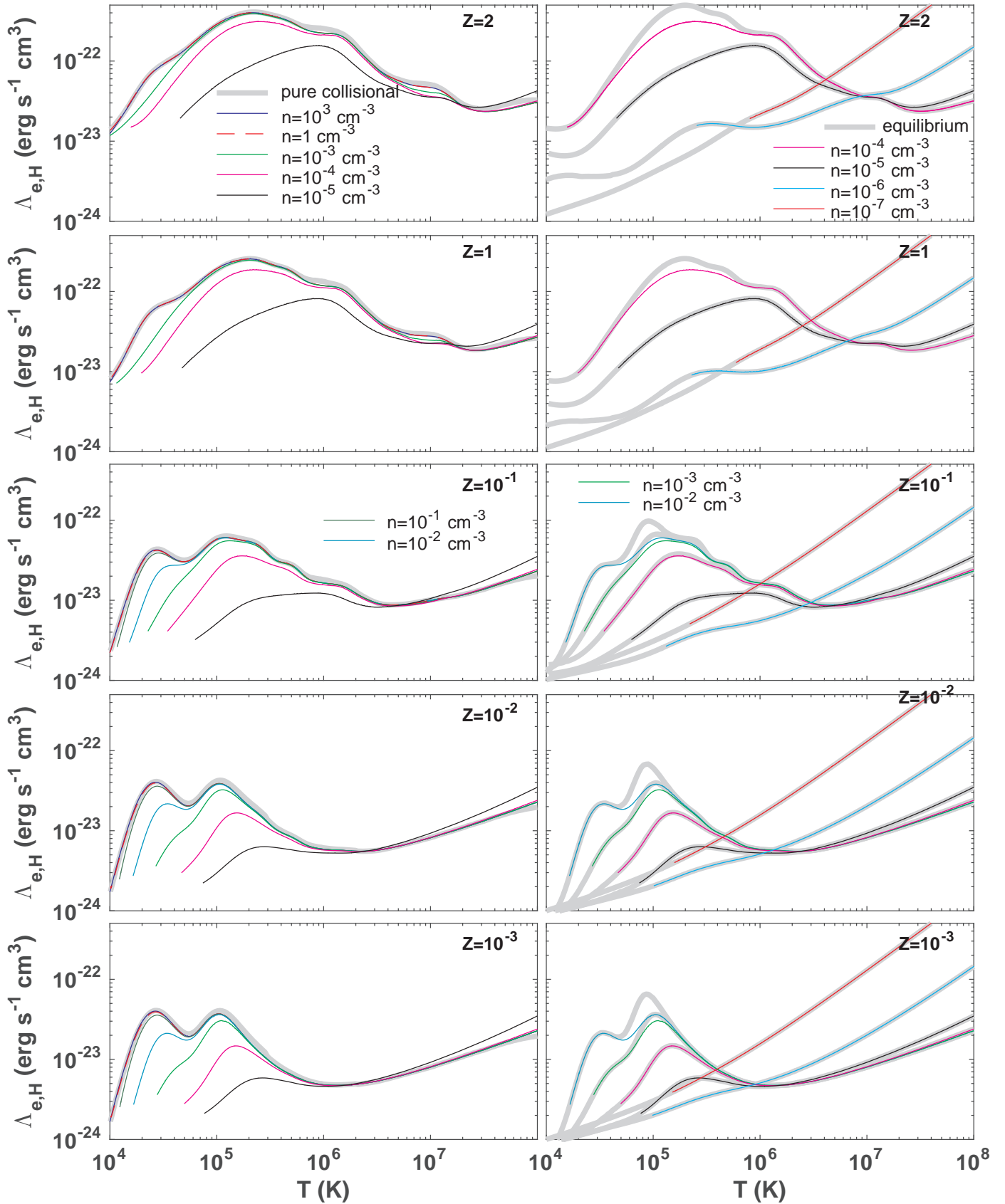}
\caption{Same as Figure~\ref{fz0}, but for the metagalactic radiation field at $z=3$.}
\label{fz3}
\end{figure*}

The photoionization-equilibrium and non-equilibrium cooling efficiencies 
for the full parameter space considered in this work are 
presented in Tables~\ref{cool-eq} and \ref{cool-neq}, respectively.
Full electronic tables are divided into lettered sections A$-$BC, as described 
in Table~\ref{Tguide}. 
Each lettered section lists the cooling efficiencies as a function of temperature
for the five metallicity values considered.
The results are displayed in Figures~\ref{fz0}-\ref{fz3}.

Focus first on the results for $z=0$, displayed in Figure~\ref{fz0}.
This figure shows the time-dependent cooling efficiencies as 
functions of temperature in gas photoionized by the $z=0$ HM12
metagalactic radiation field, for gas metallicities $Z$ equal
to $2, 1, 10^{-1}, 10^{-2}$ and $10^{-3}$ times $Z_\odot$ (top to bottom rows), 
and for gas densities between $10^3$ and $10^{-7}$~cm$^{-3}$
(shown by different colors within each row).

The non-equilibrium cooling efficiencies for ``high'' gas densities, 
$n\gtrsim10^{-4}$~cm$^{-3}$, are displayed in the left hand side panels. 
The left panels
also show the TDC (i.e. no radiation) non-equilibrium cooling
efficiency in the gray solid curve shown in the background (GS07). 
The TDC cooling efficiency (erg s$^{-1}$~cm$^{-3}$) is
independent of gas density.

In the TDC case --- shown by the thick gray curve ---
cooling is dominated by 
metal-line emission over a wide range of temperatures 
for metallicities above $10^{-2}~Z_\odot$ (three upper panels).
For lower metallicities ($Z\lesssim10^{-2}~Z_\odot$, two lower panels), 
hydrogen and helium lines, which are suppressed
by metal line cooling at higher-$Z$, dominate.
At the highest temperatures, thermal bremsstrahlung emission
dominates the energy losses. The transition between line-emission
and bremsstrahlung occurs at a few $\times10^7$~K for solar metallicity
gas, and at $\sim10^6$~K for the primordial case.

Figure \ref{fz0} shows that the various time-dependent photoionized (hereafter
TDP, see Table~\ref{glossary}) cooling curves overlap with the 
TDC curve for high densities,
where the ionization parameter is low, and the impact of 
photoionization therefore remains limited: the
solid blue, dashed red, and thick gray curves overlap.
For  densities  $\gtrsim1$~cm$^{-3}$,
the TDP cooling efficiencies are
identical to the TDC cooling efficiencies for all
gas metallicities.

Below this density, a metallicity dependence emerges.
For solar (or higher) metallicity, the TDP cooling efficiencies 
overlaps with the TDC cooling down to densities of $\sim10^{-3}$~cm$^{-3}$,
as is shown by the upper two panels.
At lower metallicities ($Z\lesssim0.1$), the cooling efficiencies in 
$10^{-3}$~cm$^{-3}$ gas overlap with the collisional curve only for 
$T\gtrsim2\times10^4$~K.
Below this temperature, cooling is suppressed, as the abundance of
neutral hydrogen is reduced by the photoionizing radiation, leading
to less efficient Ly$\alpha$ cooling.
This does not affect higher-metallicity gas, because for $Z\gtrsim1$
hydrogen remain almost completely ionized down to $10^4$~K even
with no photoionization. Metal emission lines then dominate the 
cooling down to $10^4$~K.

For even lower densities $n\le10^{-4}$~cm$^{-3}$ the photoionizing 
radiation
modifies the ion fractions of the dominant metal coolants, and
TDP cooling curves start to deviate from the 
collisional curve. For example, for $n=10^{-4}$~cm$^{-3}$ and 
$Z=1~Z_\odot$ the TDP cooling efficiencies overlap with the TDC
case for $T\gtrsim5\times10^4$~K, whereas for $Z=10^{-3}~Z_\odot$
the curves overlap only for $T\gtrsim2\times10^5$~K .

The cooling efficiencies for ``low'' gas densities,  
$\lesssim10^{-4}$~cm$^{-3}$,
are shown in the right hand side panels. 
Again, the different rows are for different gas metallicities
from $2~Z_\odot$ (top panel) to $10^{-3}~Z_\odot$ (bottom panel).
In each panel, the colored curves show the TDP cooling efficiencies
at different gas densities.
Here, the thick gray curves in the background
show the cooling efficiencies assuming {\it photoionization 
equilibrium} with the background radiation. The PIE results
{\it do} depend on the gas density, and there is 
a series of gray curves for the different densities displayed. 
The colored curves end at various temperatures, corresponding to
the thermal equilibrium temperatures at the various densities.

For densities $\gtrsim10^{-5}$~cm$^{-3}$, the cooling is 
dominated by line emission at lower temperatures, and by
thermal bremsstrahlung emission at the higher temperatures,
just as for the higher densities displayed in the left hand
side panels. However, at the lowest densities 
($n\lesssim10^{-6}$~cm$^{-3}$)
an additional cooling process comes into play, namely
Compton cooling off the metagalactic background
radiation. Compton cooling gives rise to 
the sharp increase in $\Lambda_{e,H}$ at high temperatures. 
The cooling efficiency becomes significantly higher
than the thermal bremsstrahlung efficiency.
Because Compton cooling is proportional to the electron density times 
photon density 
(see equation~\ref{compton}), as opposed to all other cooling processes 
which are proportional to the electron density squared, it dominates at low 
densities, where all other processes are suppressed. Note that 
because $\Lambda$ is defined as the cooling rate per volume divided by
$n_en_H$, the compton cooling efficiency, 
$\Lambda^{\rm compton}_{e,H} \propto n_{\rm H}^{-1}$. 

The right-hand-side panels show that for $n\le10^{-6}$ the 
TDP cooling curves overlap with the PIE
curves for all gas metallicities.
At higher gas densities departures from photoionization equilibrium
occur. For example, for $n=10^{-5}$~cm$^{-3}$ the TDP cooling curves
are identical to those in PIE for $Z<1~Z_\odot$,
whereas for $Z\ge1~Z_\odot$, slight deviations between the TDP curves
and the PIE curves are apparent near the peak of the cooling
curve, $\sim2\times10^5$~K. 

An important conclusion from Figure~\ref{fz0} is that for every
metallicity there exists
a threshold density (or, equivalently, a threshold ionization
parameter) below which the cooling efficiencies are close to 
those in PIE, and above which they
resemble those is TDC cooling gas with no external 
radiation.
Above the threshold density, departures from equilibrium ionization
are significant, but the impact of the photoionizing radiation
may be neglected.
Below this critical density, photoionization plays a significant role,
but departures from equilibrium may be neglected.

For example, for $Z=2~Z_\odot$ (top panels), the cooling efficiencies for 
$n>10^{-5}$~cm$^{-3}$ overlap with the TDC cooling efficiency,
whereas for  $n<10^{-5}$~cm$^{-3}$ the cooling curves
overlap with the PIE curves.
The threshold density, $n_{\rm th}(Z=2,z=0)=10^{-5}$~cm$^{-3}$, 
corresponds to an ionization parameter $U_{\rm th}=0.42$.
For $Z=10^{-3}~Z_\odot$ (bottom panels) the TDP cooling efficiencies for 
$n>10^{-3}$~cm$^{-3}$ overlap with the TDC cooling efficiency,
whereas for  $n<10^{-3}$~cm$^{-3}$ the TDP cooling
overlaps with the PIE curves.
The threshold density, $n_{\rm th}(Z=10^{-3},z=0)=10^{-3}$~cm$^{-3}$, 
corresponds to an ionization parameter $U_{\rm th}=4.2\times10^{-3}$.
The threshold densities/ionization parameters for $z=0$ are
listed in table~\ref{th0}.

\begin{deluxetable}{lcc}
\tablewidth{0pt}
\tablecaption{Threshold Ionization Parameters}
\tablehead{
\colhead{Metallicity} & 
\colhead{$n_{\rm th}(z=0)$}&
\colhead{$U_{\rm th}$} \\
\colhead{($Z_\odot$)}&
\colhead{(cm$^{-3}$)} &
\colhead{} }
\startdata
$1-2$           & $10^{-5}$ & $4.2\times10^{-1}$ \\
$10^{-1}-10^{-3}$ & $10^{-3}$ & $4.2\times10^{-3}$ \\
\enddata
\label{th0}
\end{deluxetable}

This finding offers a great computational simplification:
instead of following the time dependent cooling efficiencies
in the presence of photoionizing radiation, it is possible to use
pre-computed tables for the TDC cooling efficiency (i.e. GS07)
and/or for PIE.

Within $\sim$a dex around the threshold ionization parameter, both 
departures from equilibrium ionization and photoionization affect the
cooling efficiencies. However, even in this case it is enough to know
the TDC and PIE cooling efficiencies:
The TDP cooling efficiency deviates from the {\it minimum} of the
TDC and PIE efficiencies by at most $20\%$
($30\%$) for $Z\lesssim0.1~Z_\odot$ ($\gtrsim1~Z_\odot$)
(for $z=0$). 

The cooling efficiencies for redshifts $0.5,~1,~2$, and $3$
are displayed in Figures~\ref{fz05}-\ref{fz3}, and listed in
Tables~\ref{cool-eq} and~\ref{cool-neq}. As in Figure~\ref{fz0},
different rows are for different gas metallicities, and the 
different curves within each row represent different gas densities.
``High'' densities are on the left-hand side panels, and ``low'' 
densities are on the right hand side panels.
Figures~\ref{fz05}-\ref{fz3} confirm that for $z=0.5-3$ there
also exists a threshold density above which the TDP cooling
is similar to the TDC curve and below which it is
similar to that in PIE.

Figure~\ref{fz0} shows that the threshold ionization parameter
corresponding to $Z\gtrsim1~Z_\odot$ is $U=0.42$.
For $z=0$, this corresponds to a critical density of $10^{-5}$~cm$^{-3}$.
For higher redshifts, the density corresponding to this ionization parameter 
is modified as the spectral energy distribution of the metagalactic background 
evolves. The threshold densities corresponding to the threshold
ionization parameter, $U=0.42$ are displayed in Figure~\ref{ncrit}.
And indeed, Figures~\ref{fz05}-\ref{fz3} verify that these threshold densities
separate a high-density limit in which TDP cooling is equivalent to TDC cooling,
from a low-density limit in which the TDP cooling is identical to that 
in PIE.
For metallicities $\lesssim10^{-1}~Z_\odot$ the threshold ionization parameter
is a factor of $100$ lower than for solar metallicity ($U=4.2\times10^{-3}$), 
and the threshold densities are therefore a factor of $100$ higher than those 
displayed in Figure~\ref{ncrit}.

Heating in photoionized plasma is due to photoionization of atomic
and ionic species. The heating rates therefore depend on the specific 
spectral energy distribution impinging upon the gas. 
The heating rates are coupled to the ionization states in
the gas, which are affected by the same background.
Over most of the temperature range considered here,
heating is negligible with respect to cooling, and has little
impact on the thermal evolution of the gas. Hetaing only modifies 
the thermal evolution near the thermal equilibrium temperature.
For completeness, I list in Tables~\ref{heat-eq} and \ref{heat-neq}
the equilibrium and non-equilibrium heating rates for gas cooling
in the presence of the metagalactic 
background radiation at redshifts, $z=0,~0.5,~1,~2,$ and $3$ 
(HM12). A guide is provided in Table~\ref{Tguide}. 
As mentioned before, throughout this paper, the Haardt \& Madau (2012)
background radiation has been truncated at an energy of $\sim8,330$~keV. 
While this high-energy truncation point has a negligible effect of the cooling 
efficiencies, which are the main focus of this paper,
it plays a potentially important role in setting the 
absolute heating rates.  Nevertheless, it is not expected to significantly
affect the thermal evolution of low density plasmas.
This point is further discussed in Appendix~\ref{xrayapp}.

\subsection{Isobaric Cooling}

Depending on the ratio between the cooling time and the dynamical time,
cooling may proceed either isochorically (when cooling is rapid), or
isobarically (when cooling is slow). The dynamical evolution is therefore
determined by the ratio
\begin{equation}
\frac{t_c}{t_d} \propto \frac{T^{3/2}}{n\Lambda D}
\end{equation}
where $t_c\propto T/n/\Lambda$ is the cooling time, 
and $t_d\propto D/c_s \propto D/\sqrt{T}$ is the dynamical time in
a gas cloud of size $D$, and sound speed $c_s$ (see GS07 for a detailed
discussion).

For isochoric cooling, $D$ and $n$ remain constant with time. 
As the gas cools, $t_c/t_d$ generally decreases, implying that cooling which
is initially isochoric remains isochoric. For isobaric cooling, the gas contracts
as it cools. After cooling to a temperature $T$, the density will increase
by a factor $T_0/T$, and the diameter will decrease by a factor
$(T/T_0)^{1/3}$ relative to the properties at the initial temperature, $T_0$.
Thus, as the cloud cools and contracts, 
$t_c/t_d$ decreases and a transition to isochoric cooling must eventually 
occur.

\begin{deluxetable*}{lccccc}
\tablewidth{0pt}
\tablecaption{Equilibrium Heating Rates (erg s$^{-1}$)}
\tablehead{
\colhead{Temperature (K)}&
\colhead{$\Upsilon_{\rm H}(Z=10^{-3})$ }&
\colhead{$\Upsilon_{\rm H}(Z=10^{-2})$ }&
\colhead{$\Upsilon_{\rm H}(Z=10^{-1})$ }&
\colhead{$\Upsilon_{\rm H}(Z=1)$ }&
\colhead{$\Upsilon_{\rm H}(Z=2)$ }}
\startdata
$1.0000\times10^{+8}$ & $3.13\times10^{-31}$ & $3.13\times10^{-31}$ & $3.15\times10^{-31}$ & $3.38\times10^{-31}$ & $3.63\times10^{-31}$\\
$9.5010\times10^{+7}$ & $3.13\times10^{-31}$ & $3.13\times10^{-31}$ & $3.15\times10^{-31}$ & $3.39\times10^{-31}$ & $3.65\times10^{-31}$\\
$9.0270\times10^{+7}$ & $3.13\times10^{-31}$ & $3.13\times10^{-31}$ & $3.19\times10^{-31}$ & $3.40\times10^{-31}$ & $3.67\times10^{-31}$\\
$8.5770\times10^{+7}$ & $3.13\times10^{-31}$ & $3.13\times10^{-31}$ & $3.19\times10^{-31}$ & $3.41\times10^{-31}$ & $3.70\times10^{-31}$\\
$\vdots$  &  $\vdots$ &  $\vdots$ &  $\vdots$ &  $\vdots$ &  $\vdots$ \\
\enddata
\tablecomments{The complete version of this table is in 
the electronic edition of the Journal. The printed edition 
contains only a sample. 
The full table is divided into lettered parts A$-$BC, and
lists the equilibrium heating rates
for for gas densities between $10^{-7}$ and $10^3$~cm$^{-3}$, for 
the metagalactic backgrounds at redshifts $0-3$, and for
$Z=10^{-3}$, $10^{-2}$, $10^{-1}$, $1$, and $2$ times solar metallicity
gas (for a guide, see Table~\ref{Tguide}).}
\label{heat-eq}
\end{deluxetable*}

\begin{deluxetable*}{lccccc}
\tablewidth{0pt}
\tablecaption{Non-equilibrium Heating Rates (erg s$^{-1}$)}
\tablehead{
\colhead{Temperature (K)}&
\colhead{$\Upsilon_{\rm H}(Z=10^{-3})$ }&
\colhead{$\Upsilon_{\rm H}(Z=10^{-2})$ }&
\colhead{$\Upsilon_{\rm H}(Z=10^{-1})$ }&
\colhead{$\Upsilon_{\rm H}(Z=1)$ }&
\colhead{$\Upsilon_{\rm H}(Z=2)$ }}
\startdata
$1.0000\times10^{+8}$ & $3.13\times10^{-31}$ & $3.13\times10^{-31}$ & $3.15\times10^{-31}$ & $3.37\times10^{-31}$ & $3.62\times10^{-31}$\\
$9.9000\times10^{+7}$ & $3.13\times10^{-31}$ & $3.13\times10^{-31}$ & $3.15\times10^{-31}$ & $3.38\times10^{-31}$ & $3.63\times10^{-31}$\\
$9.8010\times10^{+7}$ & $3.13\times10^{-31}$ & $3.13\times10^{-31}$ & $3.15\times10^{-31}$ & $3.38\times10^{-31}$ & $3.64\times10^{-31}$\\
$9.7030\times10^{+7}$ & $3.13\times10^{-31}$ & $3.13\times10^{-31}$ & $3.15\times10^{-31}$ & $3.39\times10^{-31}$ & $3.64\times10^{-31}$\\
$\vdots$  &  $\vdots$ &  $\vdots$ &  $\vdots$ &  $\vdots$ &  $\vdots$ \\
\enddata
\tablecomments{The complete version of this table is in 
the electronic edition of the Journal. The printed edition 
contains only a sample. 
The full table is divided into lettered parts A$-$BC, and
lists the non-equilibrium heating rates
for for gas densities between $10^{-7}$ and $10^3$~cm$^{-3}$, for 
the metagalactic backgrounds at redshifts $0-3$, and for
$Z=10^{-3}$, $10^{-2}$, $10^{-1}$, $1$, and $2$ times solar metallicity
gas (for a guide, see Table~\ref{Tguide}).}
\label{heat-neq}
\end{deluxetable*}

When the gas cools isobarically in the presence of photoionizing radiation 
the varying density also implies a varying ionization parameter.
Consider the evolution of an isobarically cooling gas cloud.
A gas cloud that begins the cooling process at a density larger than
the threshold density, will always remain at a higher density.
In this regime, one may consider the TDC cooling efficiency 
throughout the cooling process.

For a gas cloud that is initially at a density lower than the threshold
density, cooling will initially proceed according to the PIE efficiencies.
However, the density will increase by a factor ($T_0/T$) after cooling 
to a temperature T. The cloud may cross the threshold density,
thus transitioning from the regime in which the PIE cooling efficiencies
apply, into the regime where the TDC cooling efficiency apply.

\begin{figure}
\vspace{1cm}
\epsscale{1.2}
\plotone{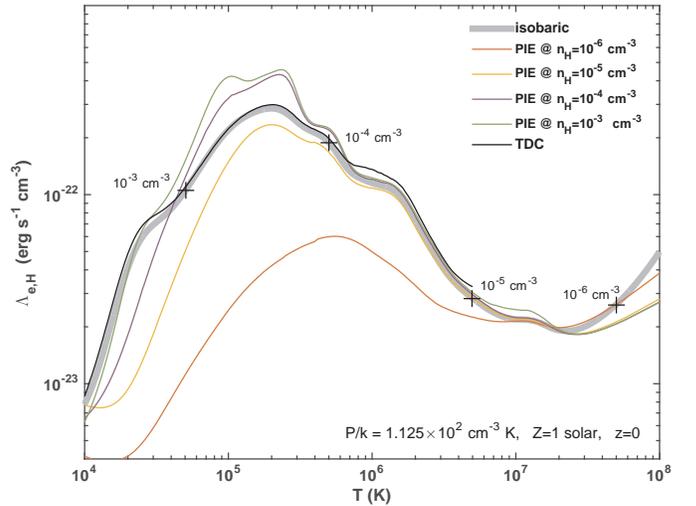}
\caption{Isobaric cooling efficiency (thick gray) for gas exposed to the $z=0$
metagalactic radiation field,
for an assumed pressure of 112.5~cm$^{-3}$~K, and for solar abundances.
The colored curves show the PIE cooling efficiencies for solar metallicity gas
at densities between $10^{-6}$ and $10^{-3}$~cm$^{-3}$ (red to green).
The black curve shows the isobaric TDC cooling efficiency.
The crosses and labels mark the hydrogen density at certain points along the
cooling process. The time-dependent isobaric cooling matches the PIE cooling
at the appropriate local density for densities below the threshold density,
and the TDC cooling efficiency for densities greater than the threshold density.}
\label{isobaric}
\end{figure}

This point in demonstrated in Figure~\ref{isobaric}. The thick gray curve
depicts the time-dependent cooling efficiency in an {\it isobarically} cooling
gas cloud with a constant pressure $P=112.5$~cm$^{-3}$~K, and solar 
metallicity, 
exposed to the $z=0$ metagalactic radiation field.
This gas begins its cooling process at an initial temperature $T_0=10^8$~K,
and density $n_{{\rm H},0}=5\times10^{-7}$~cm$^{-3}$, and achieves
thermal equilibrium at a temperature $T_{\rm eq}=1.13\times10^4$~K,
when its density is $n_{{\rm}eq}=4.8\times10^{-3}$~cm$^{-3}$.
The crosses and labels mark the hydrogen density at certain points along the
cooling process.

The colored curves displayed in Figure~\ref{isobaric} show the PIE
cooling efficiencies at densities of $10^{-6} - 10^{-3}$~cm$^{-3}$
(from red to green). The black curve shows the (isobaric) TDC cooling efficiency (GS07).
The cooling efficiency is indeed identical to the PIE cooling efficiency
at the appropriate density, as long as the gas density is lower
than the threshold density.
For example, the isobaric thick gray curve and the PIE red curve intersect
just when the density equals $10^{-6}$~cm$^{-3}$ (marked by the cross).
The isobaric curve crosses the orange curve when the density
equals $10^{-5}$~cm$^{-3}$, as expected.
At $\sim10^{-5}$~cm$^{-3}$ the gas crosses the threshold density,
and at lower temperatures (higher densities) we therefore expect the
cooling efficiency to match the TDC cooling efficiency.
And indeed, at lower temperature, the isobaric gray curve matches the TDC
black curve, and not the (purple, green) PIE curves.

Figure~\ref{isobaric} demonstrates that the threshold density can be used
to separate the regime in which PIE cooling efficiencies apply from
the regime in which the cooling efficiency is the TDC efficiency, even
for the case of isobaric cooling.

\section{Ion Fractions}
\label{ionization}

I have carried out computations of the photoionized ion fractions 
of the elements H, He, C, N, O, Ne, Mg, Si, S, and Fe, as functions 
of gas temperature for two sets of assumptions. First I assume PIE
imposed at all T. Then I consider the nonequilibrium ion fractions
in TDP gas as a function of the time-dependent temperature.
The results are listed in tabular form in Tables~\ref{eqion} and~\ref{tdion}
for the PIE and TDP ion fractions, respectively.
These tables are divided into numbered parts $1-275$, as described in Table~\ref{ionguide}.

In \S\ref{cooling}, I introduced the threshold ionization parameter,
which separates the regime in which radiation significantly affects
the cooling efficiencies (but departures from equilibrium may be neglected), from the regime
in which non-equilibrium effects become important (but radiation may be 
neglected). Here I consider the corresponding behavior of the ion fractions
for different gas densities.

\begin{figure}
\vspace{0.5cm}
\epsscale{1.2}
\plotone{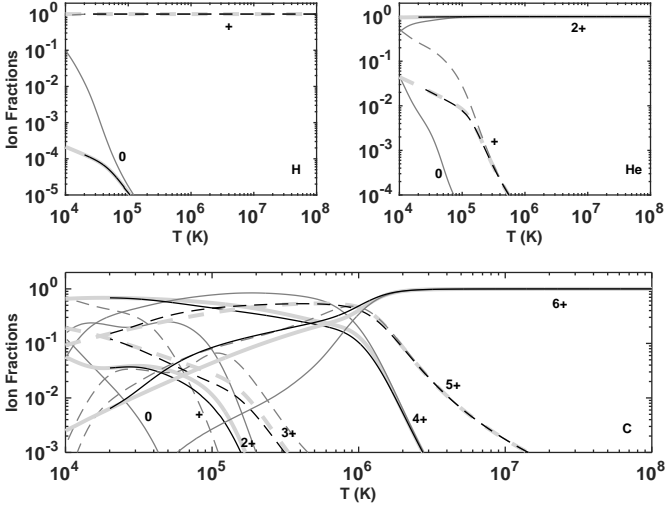}
\caption{Ion fractions $x_i ≡ n_{i,m}/n_H A_m$, vs. gas temperature, for $n_H=10^{-5}$~cm$^{-3}$, 
solar metallicity gas exposed to the $z=0$ metagalactic radiation. The TDP ion fractions are 
displayed by the dark curves, the PIE ion fractions are in 
thick gray curves, and the TDC ion fractions are in the thin gray curves.
The different ionization states are shown by the dashed and solid curves (alternately;
Neutrals are in solid, singly ionized dashed, etc.)
and are labeled in each panel. The top panels are for hydrogen (left), and helium (right)
and the bottom panel is for carbon.}
\label{n5ion}
\end{figure}

For solar metallicity gas exposed to the $z=0$ metagalactic background,
the threshold ionization parameter corresponds to a threshold hydrogen 
density $n_H = 10^{-5}$~cm$^{-3}$.
Figure~\ref{n5ion} presents the evolution of some ion fractions
for this threshold gas density (for $Z=1 Z_\odot$, and the $z=0$
metagalactic background). This is the density for which the 
impact of time-dependent photoionization should be maximal.
In Figure~\ref{n5ion}, the dark curves present the results for the 
TDP ion fractions. This is compared to the ion fractions in PIE
(thick gray curves), as well as to the ion fractions in 
TDC cooling gas (thin gray curves). The top panels display
the ion fractions of hydrogen (left) and helium (right) as functions
of temperature, and the bottom panel is for carbon.
The different ionization states in each panel are denoted by dashed 
and solid curves (alternately), and are labeled next to the curves.

Figure~\ref{n5ion} shows that for this density gas in PIE (thick gray)
is more highly ionized than gas overionized due to time-dependent 
cooling in the absence of external radiation (TDC, thin gray).
Moreover, TDP cooling leads to recombination lags which cause the gas 
to be ionized above and beyond the ionization in PIE.
For example, at a temperature of $3\times10^5$~K,
the C$^{3+}$, C$^{4+}$, C$^{5+}$, and C$^{6+}$ fractions are
$3.9\times10^{-3}$, $8.0\times10^{-1}$, $1.8\times10^{-1}$, and $1.3\times10^{-2}$ in TDC gas,
$1.6\times10^{3}$,  $3.6\times10^{-1}$, $5.1\times10^{-1}$, and $1.3\times10^{-1}$ for PIE, and 
$1.2\times10^{-3}$, $2.9\times10^{-1}$, $5.3\times10^{-1}$, and $1.8\times10^{1}$
in TDP cooling gas.
As far as the ionization states are concerned, the threshold density
represents a case for which the ion fractions significantly deviate 
(by factors up to $4$) from both PIE and TDC ion fractions. This is 
in contrast to the cooling efficiencies which deviate from the {\it minimum} 
of the TDC and PIE efficiencies by at most $20-30\%$ (see 
Section~\ref{cooling}). 

\begin{turnpage}
\begin{deluxetable*}{cccl|cccl|cccl|cccl|cccl}
\vspace{-3.3cm}
\tabletypesize{\scriptsize}
\renewcommand{\arraystretch}{0.52}
\tablewidth{0pt}
\tablecaption{Ionization Data Guide \label{ionguide}}
\tablehead{
\colhead{$n_{\rm H}$}&
\colhead{$z$}&
\colhead{$Z$}&
\colhead{part}&
\colhead{$n_{\rm H}$}&
\colhead{$z$}&
\colhead{$Z$}&
\colhead{part}&
\colhead{$n_{\rm H}$}&
\colhead{$z$}&
\colhead{$Z$}&
\colhead{part}&
\colhead{$n_{\rm H}$}&
\colhead{$z$}&
\colhead{$Z$}&
\colhead{part}&
\colhead{$n_{\rm H}$}&
\colhead{$z$}&
\colhead{$Z$}&
\colhead{part}\\
\colhead{$($cm$^{-3})$}&
\colhead{}&
\colhead{$(Z_\odot)$}&
\colhead{}&
\colhead{$($cm$^{-3})$}&
\colhead{}&
\colhead{$(Z_\odot)$}&
\colhead{}&
\colhead{$($cm$^{-3})$}&
\colhead{}&
\colhead{$(Z_\odot)$}&
\colhead{}&
\colhead{$($cm$^{-3})$}&
\colhead{}&
\colhead{$(Z_\odot)$}&
\colhead{}&
\colhead{$($cm$^{-3})$}&
\colhead{}&
\colhead{$(Z_\odot)$}&
\colhead{} }
\startdata
$10^{-7}$&$3$  &$10^{-3}$&$  1$&$10^{-5}$&$2$  &$10^{-3}$&$ 56$&$10^{-3}$&$1$  &$10^{-3}$&$111$& $10^{-1}$&$0.5$&$10^{-3}$&$166$& $10     $&$0$  &$10^{-3}$&$221$\\
$10^{-7}$&$3$  &$10^{-2}$&$  2$&$10^{-5}$&$2$  &$10^{-2}$&$ 57$&$10^{-3}$&$1$  &$10^{-2}$&$112$& $10^{-1}$&$0.5$&$10^{-2}$&$167$& $10     $&$0$  &$10^{-2}$&$222$\\
$10^{-7}$&$3$  &$10^{-1}$&$  3$&$10^{-5}$&$2$  &$10^{-1}$&$ 58$&$10^{-3}$&$1$  &$10^{-1}$&$113$& $10^{-1}$&$0.5$&$10^{1}$&$168$& $10      $&$0$  &$10^{-1}$&$223$\\
$10^{-7}$&$3$  &$1     $&$  4$&$10^{-5}$&$2$  &$1     $&$ 59$& $10^{-3}$&$1$  &$1     $&$114$& $10^{-1}$&$0.5$&$1     $&$169$& $10      $&$0$  &$1     $&$224$\\
$10^{-7}$&$3$  &$2     $&$  5$&$10^{-5}$&$2$  &$2     $&$ 60$& $10^{-3}$&$1$  &$2     $&$115$& $10^{-1}$&$0.5$&$2     $&$170$& $10      $&$0$  &$2     $&$225$\\
$10^{-7}$&$2$  &$10^{-3}$&$  6$&$10^{-5}$&$1$  &$10^{-3}$&$ 61$&$10^{-3}$&$0.5$&$10^{-3}$&$116$& $10^{-1}$&$0$  &$10^{-3}$&$171$& $10^{2 }$&$3$  &$10^{-3}$&$226$\\
$10^{-7}$&$2$  &$10^{-2}$&$  7$&$10^{-5}$&$1$  &$10^{-2}$&$ 62$&$10^{-3}$&$0.5$&$10^{-2}$&$117$& $10^{-1}$&$0$  &$10^{-2}$&$172$& $10^{2 }$&$3$  &$10^{-2}$&$227$\\
$10^{-7}$&$2$  &$10^{-1}$&$  8$&$10^{-5}$&$1$  &$10^{-3}$&$ 63$&$10^{-3}$&$0.5$&$10^{-1}$&$118$& $10^{-1}$&$0$  &$10^{-1}$&$173$& $10^{2 }$&$3$  &$10^{-1}$&$228$\\
$10^{-7}$&$2$  &$1     $&$  9$&$10^{-5}$&$1$  &$1     $&$ 64$& $10^{-3}$&$0.5$&$1     $&$119$& $10^{-1}$&$0$  &$1     $&$174$&  $10^{2 }$&$3$  &$1     $&$229$\\
$10^{-7}$&$2$  &$2     $&$ 10$&$10^{-5}$&$1$  &$2     $&$ 65$& $10^{-3}$&$0.5$&$2     $&$120$& $10^{-1}$&$0$  &$2     $&$175$&  $10^{2 }$&$3$  &$2     $&$230$\\
$10^{-7}$&$1$  &$10^{-3}$&$ 11$&$10^{-5}$&$0.5$&$10^{-3}$&$ 66$&$10^{-3}$&$0$  &$10^{-3}$&$121$& $1      $&$3$  &$10^{-3}$&$176$&  $10^{2 }$&$2$  &$10^{-3}$&$231$\\
$10^{-7}$&$1$  &$10^{-2}$&$ 12$&$10^{-5}$&$0.5$&$10^{-2}$&$ 67$&$10^{-3}$&$0$  &$10^{-2}$&$122$& $1      $&$3$  &$10^{-2}$&$177$& $10^{2 }$&$2$  &$10^{-2}$&$232$\\
$10^{-7}$&$1$  &$10^{-1}$&$ 13$&$10^{-5}$&$0.5$&$10^{-1}$&$ 68$&$10^{-3}$&$0$  &$10^{-1}$&$123$& $1      $&$3$  &$10^{-1}$&$178$& $10^{2 }$&$2$  &$10^{-1}$&$233$\\
$10^{-7}$&$1$  &$1     $&$ 14$&$10^{-5}$&$0.5$&$1     $&$ 69$& $10^{-3}$&$0$  &$1     $&$124$& $1      $&$3$  &$1      $&$179$& $10^{2 }$&$2$  &$1     $&$234$\\
$10^{-7}$&$1$  &$2     $&$ 15$&$10^{-5}$&$0.5$&$2     $&$ 70$& $10^{-3}$&$0$  &$2     $&$125$& $1      $&$3$  &$2      $&$180$& $10^{2 }$&$2$  &$2     $&$235$\\
$10^{-7}$&$0.5$&$10^{-3}$&$ 16$&$10^{-5}$&$0$  &$10^{-3}$&$ 71$&$10^{-2}$&$3$  &$10^{-3}$&$126$& $1      $&$2$  &$10^{-3}$&$181$& $10^{2 }$&$1$  &$10^{-3}$&$236$\\
$10^{-7}$&$0.5$&$10^{-2}$&$ 17$&$10^{-5}$&$0$  &$10^{-2}$&$ 72$&$10^{-2}$&$3$  &$10^{-2}$&$127$& $1      $&$2$  &$10^{-2}$&$182$& $10^{2 }$&$1$  &$10^{-2}$&$237$\\
$10^{-7}$&$0.5$&$10^{-1}$&$ 18$&$10^{-5}$&$0$  &$10^{-1}$&$ 73$&$10^{-2}$&$3$  &$10^{-1}$&$128$& $1      $&$2$  &$10^{-1}$&$183$& $10^{2 }$&$1$  &$10^{-1}$&$238$\\
$10^{-7}$&$0.5$&$1     $&$ 19$&$10^{-5}$&$0$  &$1     $&$ 74$& $10^{-2}$&$3$  &$1     $&$129$& $1      $&$2$  &$1      $&$184$& $10^{2 }$&$1$  &$1     $&$239$\\
$10^{-7}$&$0.5$&$2     $&$ 20$&$10^{-5}$&$0$  &$2     $&$ 75$& $10^{-2}$&$3$  &$2     $&$130$& $1      $&$2$  &$2      $&$185$& $10^{2 }$&$1$  &$2     $&$240$\\
$10^{-7}$&$0$  &$10^{-3}$&$ 21$&$10^{-4}$&$3$  &$10^{-3}$&$ 76$&$10^{-2}$&$2$  &$10^{-3}$&$131$& $1      $&$1$  &$10^{-3}$&$186$& $10^{2 }$&$0.5$&$10^{-3}$&$241$\\
$10^{-7}$&$0$  &$10^{-2}$&$ 22$&$10^{-4}$&$3$  &$10^{-2}$&$ 77$&$10^{-2}$&$2$  &$10^{-2}$&$132$& $1      $&$1$  &$10^{-2}$&$187$& $10^{2 }$&$0.5$&$10^{-2}$&$242$\\
$10^{-7}$&$0$  &$10^{-1}$&$ 23$&$10^{-4}$&$3$  &$10^{-1}$&$ 78$&$10^{-2}$&$2$  &$10^{-1}$&$133$& $1      $&$1$  &$10^{-1}$&$188$& $10^{2 }$&$0.5$&$10^{-1}$&$243$\\
$10^{-7}$&$0$  &$1     $&$ 24$&$10^{-4}$&$3$  &$1     $&$ 79$& $10^{-2}$&$2$  &$1     $&$134$& $1      $&$1$  &$1      $&$189$& $10^{2 }$&$0.5$&$1     $&$244$\\
$10^{-7}$&$0$  &$2     $&$ 25$&$10^{-4}$&$3$  &$2     $&$ 80$& $10^{-2}$&$2$  &$2     $&$135$& $1      $&$1$  &$2      $&$190$& $10^{2 }$&$0.5$&$2     $&$245$\\
$10^{-6}$&$3$  &$10^{-3}$&$ 26$&$10^{-4}$&$2$  &$10^{-3}$&$ 81$&$10^{-2}$&$1$  &$10^{-3}$&$136$& $1      $&$0.5$&$10^{-3}$&$191$& $10^{2 }$&$0$  &$10^{-3}$&$246$\\
$10^{-6}$&$3$  &$10^{-2}$&$ 27$&$10^{-4}$&$2$  &$10^{-2}$&$ 82$&$10^{-2}$&$1$  &$10^{-2}$&$137$& $1      $&$0.5$&$10^{-2}$&$192$& $10^{2 }$&$0$  &$10^{-2}$&$247$\\
$10^{-6}$&$3$  &$10^{-1}$&$ 28$&$10^{-4}$&$2$  &$10^{-1}$&$ 83$&$10^{-2}$&$1$  &$10^{-1}$&$138$& $1      $&$0.5$&$10^{-1}$&$193$& $10^{2 }$&$0$  &$10^{-1}$&$248$\\
$10^{-6}$&$3$  &$1     $&$ 29$&$10^{-4}$&$2$  &$1     $&$ 84$& $10^{-2}$&$1$  &$1     $&$139$& $1      $&$0.5$&$1      $&$194$& $10^{2 }$&$0$  &$1     $&$249$\\
$10^{-6}$&$3$  &$2     $&$ 30$&$10^{-4}$&$2$  &$2     $&$ 85$& $10^{-2}$&$1$  &$2     $&$140$& $1      $&$0.5$&$2      $&$195$& $10^{2 }$&$0$  &$2     $&$250$\\
$10^{-6}$&$2$  &$10^{-3}$&$ 31$&$10^{-4}$&$1$  &$10^{-3}$&$ 86$&$10^{-2}$&$0.5$&$10^{-3}$&$141$& $1      $&$0$  &$10^{-3}$&$196$& $10^{3 }$&$3$  &$10^{-3}$&$251$\\
$10^{-6}$&$2$  &$10^{-2}$&$ 32$&$10^{-4}$&$1$  &$10^{-2}$&$ 87$&$10^{-2}$&$0.5$&$10^{-2}$&$142$& $1      $&$0$  &$10^{-2}$&$197$& $10^{3 }$&$3$  &$10^{-2}$&$252$\\
$10^{-6}$&$2$  &$10^{-1}$&$ 33$&$10^{-4}$&$1$  &$10^{-1}$&$ 88$&$10^{-2}$&$0.5$&$10^{-1}$&$143$& $1      $&$0$  &$10^{-1}$&$198$& $10^{3 }$&$3$  &$10^{-1}$&$253$\\
$10^{-6}$&$2$  &$1     $&$ 34$&$10^{-4}$&$1$  &$1     $&$ 89$& $10^{-2}$&$0.5$&$1     $&$144$& $1      $&$0$  &$1      $&$199$& $10^{3 }$&$3$  &$1      $&$254$\\
$10^{-6}$&$2$  &$2     $&$ 35$&$10^{-4}$&$1$  &$2     $&$ 90$& $10^{-2}$&$0.5$&$2     $&$145$& $1      $&$0$  &$2      $&$200$& $10^{3 }$&$3$  &$2     $&$255$\\
$10^{-6}$&$1$  &$10^{-3}$&$ 36$&$10^{-4}$&$0.5$&$10^{-3}$&$ 91$&$10^{-2}$&$0$  &$10^{-3}$&$146$& $10     $&$3$  &$10^{-3}$&$201$& $10^{3 }$&$2$  &$10^{-3}$&$256$\\
$10^{-6}$&$1$  &$10^{-2}$&$ 37$&$10^{-4}$&$0.5$&$10^{-2}$&$ 92$&$10^{-2}$&$0$  &$10^{-2}$&$147$& $10     $&$3$  &$10^{-2}$&$202$& $10^{3 }$&$2$  &$10^{-2}$&$257$\\
$10^{-6}$&$1$  &$10^{-1}$&$ 38$&$10^{-4}$&$0.5$&$10^{-1}$&$ 93$&$10^{-2}$&$0$  &$10^{-1}$&$148$& $10     $&$3$  &$10^{-1}$&$203$& $10^{3 }$&$2$  &$10^{-1}$&$258$\\
$10^{-6}$&$1$  &$1     $&$ 39$&$10^{-4}$&$0.5$&$1     $&$ 94$& $10^{-2}$&$0$  &$1     $&$149$& $10     $&$3$  &$1      $&$204$& $10^{3 }$&$2$  &$1     $&$259$\\
$10^{-6}$&$1$  &$2     $&$ 40$&$10^{-4}$&$0.5$&$2     $&$ 95$& $10^{-2}$&$0$  &$2     $&$150$& $10     $&$3$  &$2      $&$205$& $10^{3 }$&$2$  &$2     $&$260$\\
$10^{-6}$&$0.5$&$10^{-3}$&$ 41$&$10^{-4}$&$0$  &$10^{-3}$&$ 96$&$10^{-1}$&$3$  &$10^{-3}$&$151$& $10     $&$2$  &$10^{-3}$&$206$& $10^{3 }$&$1$  &$10^{-3}$&$261$\\
$10^{-6}$&$0.5$&$10^{-2}$&$ 42$&$10^{-4}$&$0$  &$10^{-2}$&$ 97$&$10^{-1}$&$3$  &$10^{-2}$&$152$& $10     $&$2$  &$10^{-2}$&$207$& $10^{3 }$&$1$  &$10^{-2}$&$262$\\
$10^{-6}$&$0.5$&$10^{-1}$&$ 43$&$10^{-4}$&$0$  &$10^{-1}$&$ 98$&$10^{-1}$&$3$  &$10^{-1}$&$153$& $10     $&$2$  &$10^{-1}$&$208$& $10^{3 }$&$1$  &$10^{-1}$&$263$\\
$10^{-6}$&$0.5$&$1     $&$ 44$&$10^{-4}$&$0$  &$1     $&$ 99$& $10^{-1}$&$3$  &$1     $&$154$& $10     $&$2$  &$1      $&$209$& $10^{3 }$&$1$  &$1     $&$264$\\
$10^{-6}$&$0.5$&$2     $&$ 45$&$10^{-4}$&$0$  &$2     $&$100$& $10^{-1}$&$3$  &$2     $&$155$& $10     $&$2$  &$2      $&$210$& $10^{3 }$&$1$  &$2     $&$265$\\
$10^{-6}$&$0$  &$10^{-3}$&$ 46$&$10^{-3}$&$3$  &$10^{-3}$&$101$&$10^{-1}$&$2$  &$10^{-3}$&$156$& $10     $&$1$  &$10^{-3}$&$211$& $10^{3 }$&$0.5$&$10^{-3}$&$266$\\
$10^{-6}$&$0$  &$10^{-2}$&$ 47$&$10^{-3}$&$3$  &$10^{-2}$&$102$&$10^{-1}$&$2$  &$10^{-2}$&$157$& $10     $&$1$  &$10^{-2}$&$212$& $10^{3 }$&$0.5$&$10^{-2}$&$267$\\
$10^{-6}$&$0$  &$10^{-1}$&$ 48$&$10^{-3}$&$3$  &$10^{-1}$&$103$&$10^{-1}$&$2$  &$10^{-1}$&$158$& $10     $&$1$  &$10^{-1}$&$213$& $10^{3 }$&$0.5$&$10^{-1}$&$268$\\
$10^{-6}$&$0$  &$1     $&$ 49$&$10^{-3}$&$3$  &$1     $&$104$& $10^{-1}$&$2$  &$1     $&$169$& $10     $&$1$  &$1      $&$214$& $10^{3 }$&$0.5$&$1     $&$269$\\
$10^{-6}$&$0$  &$2     $&$ 50$&$10^{-3}$&$3$  &$2     $&$105$& $10^{-1}$&$2$  &$2     $&$160$& $10     $&$1$  &$2      $&$215$& $10^{3 }$&$0.5$&$2     $&$270$\\
$10^{-5}$&$3$  &$10^{-3}$&$ 51$&$10^{-3}$&$2$  &$10^{-3}$&$106$&$10^{-1}$&$1$  &$10^{-3}$&$161$& $10     $&$0.5$&$10^{-3}$&$216$& $10^{3 }$&$0$  &$10^{-3}$&$271$\\
$10^{-5}$&$3$  &$10^{-2}$&$ 52$&$10^{-3}$&$2$  &$10^{-2}$&$107$&$10^{-1}$&$1$  &$10^{-2}$&$162$& $10     $&$0.5$&$10^{-2}$&$217$& $10^{3 }$&$0$  &$10^{-2}$&$272$\\
$10^{-5}$&$3$  &$10^{-1}$&$ 53$&$10^{-3}$&$2$  &$10^{-1}$&$108$&$10^{-1}$&$1$  &$10^{-1}$&$163$& $10     $&$0.5$&$10^{-1}$&$218$& $10^{3 }$&$0$  &$10^{-1}$&$273$\\
$10^{-5}$&$3$  &$1     $&$ 54$&$10^{-3}$&$2$  &$1     $&$109$& $10^{-1}$&$1$  &$1     $&$164$& $10     $&$0.5$&$1      $&$219$& $10^{3 }$&$0$  &$1     $&$274$\\
$10^{-5}$&$3$  &$2     $&$ 55$&$10^{-3}$&$2$  &$2     $&$110$& $10^{-1}$&$1$  &$2     $&$165$& $10     $&$0.5$&$2      $&$220$& $10^{3 }$&$0$  &$2     $&$275$\\
\enddata
\tablecomments{Table~\ref{eqion} (\ref{tdion}) lists the equilibrium
(time-dependent) ion fraction in photoionized gas. Tables~\ref{eqion} 
and~\ref{tdion} are divided into numerical parts $1-275$.
Each part lists the ion fraction for a specific set of density, metallicity, and
redshift, as described here.}
\end{deluxetable*}
\end{turnpage}
\global\pdfpageattr\expandafter{\the\pdfpageattr/Rotate 90}

\begin{figure}
\vspace{1cm}
\epsscale{1.2}
\plotone{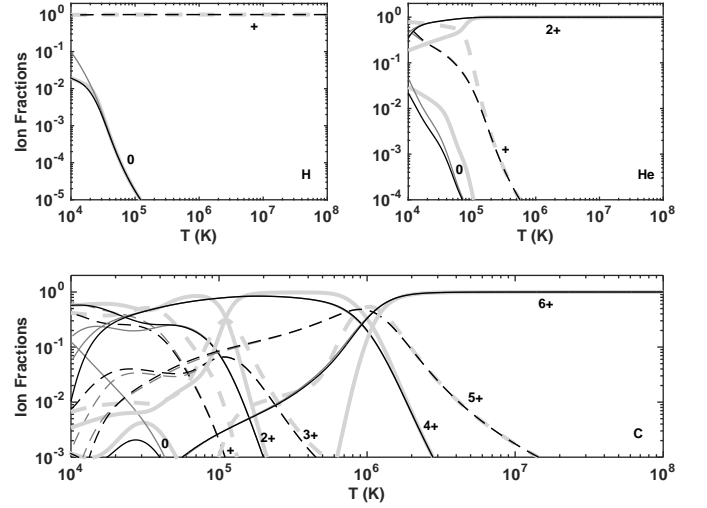}
\caption{Same as Figure~\ref{n3ion}, but for $n_H=10^{-3}$~cm$^{-3}$}
\label{n3ion}
\end{figure}

Figure~\ref{n3ion} displays the ion fractions for gas $2$ dex above 
the threshold density, $n_H=10^{-3}$~cm$^{-3}$. For this density, the TDP
cooling efficiencies are identical to the TDC cooling efficiencies
for all temperatures above $4\times10^4$~K, and even at for lower 
temperatures the differences remain limited.
However, the abundance of certain species (e.g. Si$^{5+}$) are more 
strongly affected by the recombination lags in the photoionized 
cooling gas, and may differ significantly (by factors up to $\sim4$)
from those of the TDC cooling gas even at higher temperatures.

\begin{figure}
\epsscale{1.2}
\plotone{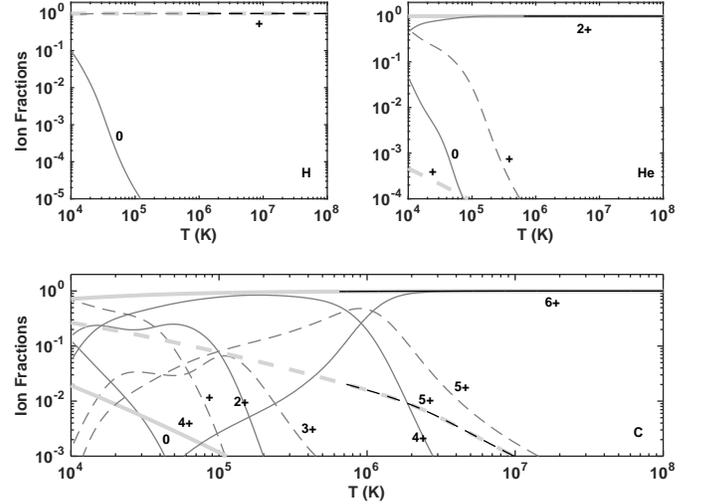}
\caption{Same as Figure~\ref{n3ion}, but for $n_H=10^{-7}$~cm$^{-3}$}
\label{n7ion}
\end{figure}

Figure~\ref{n7ion} considers a hydrogen density of $10^{-7}$~cm$^{-3}$,
$2$ dex below the threshold density. In this case, the cooling gas
reaches thermal equilibrium at a temperature $\sim7\times10^5$~K.
Throughout the cooling process, all ion fractions are identical
(to within a few $\%$) to those in PIE.

\begin{deluxetable}{ccccc}
\tablewidth{0pt}
\tablecaption{Photoionization Equilibrium Ion Fractions\label{eqion}}
\tablehead{
\colhead{Temperature}&
\colhead{}&
\colhead{}&
\colhead{}&
\colhead{}\\
\colhead{(K)}&
\colhead{H$^0/$H}&
\colhead{H$^+/$H}&
\colhead{He$^0/$He}&
\colhead{$\dots$}
}
\startdata
$1.00\times10^{8}$& $0.00$& $1.00$& $0.00$& $\dots$\\ 
$9.50\times10^{7}$& $0.00$& $1.00$& $0.00$& $\dots$\\ 
$9.03\times10^{7}$& $0.00$& $1.00$& $0.00$& $\dots$\\ 
$\vdots$       &$\vdots$&$\vdots$&$\vdots$&$\vdots$\\
\enddata
\tablecomments{Table~\ref{eqion} is listed in its entirety in the electronic 
edition of the {\it Astrophysical Journal Supplement}. A portion is shown 
here for guidance. This table is divided into $275$ numerical parts, each
corresponding to specific density, redshift, and metallicity as
described in Table~\ref{ionguide}.}
\end{deluxetable}

\begin{deluxetable}{ccccc}
\tablewidth{0pt}
\tablecaption{Time-Dependent Photoionized Ion Fractions\label{tdion}}
\tablehead{
\colhead{Temperature}&
\colhead{}&
\colhead{}&
\colhead{}&
\colhead{}\\
\colhead{(K)}&
\colhead{H$^0/$H}&
\colhead{H$^+/$H}&
\colhead{He$^0/$He}&
\colhead{$\dots$}
}
\startdata
$1.00\times10^{8}$& $0.00$& $1.00$& $0.00$& $\dots$\\ 
$9.90\times10^{7}$& $0.00$& $1.00$& $0.00$& $\dots$\\ 
$9.80\times10^{7}$& $0.00$& $1.00$& $0.00$& $\dots$\\ 
$\vdots$       &$\vdots$&$\vdots$&$\vdots$&$\vdots$\\
\enddata
\tablecomments{Table~\ref{tdion} is listed in its entirety in the electronic 
edition of the {\it Astrophysical Journal Supplement}. A portion is shown 
here for guidance. This table is divided into $275$ numerical parts, each
corresponding to specific density, redshift, and metallicity as
described in Table~\ref{ionguide}.}
\end{deluxetable}


\begin{figure}
\epsscale{1.2}
\plotone{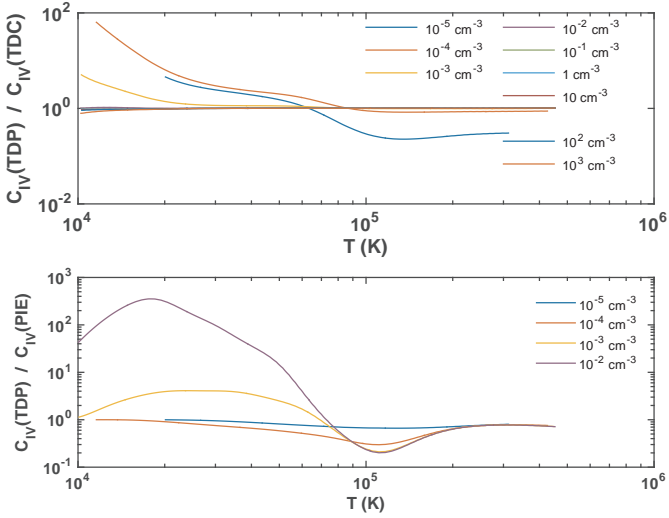}
\caption{Ratio of C$^{3+}$ TDP ion fraction and TDC ion fraction versus 
temperature (top panel), and ratio between C$^{3+}$ TDP ion fraction and 
PIE ion fraction versus temperature (bottom panels). Different curves 
are for different densities.}
\label{crat}
\end{figure}

\begin{figure}
\epsscale{1.2}
\plotone{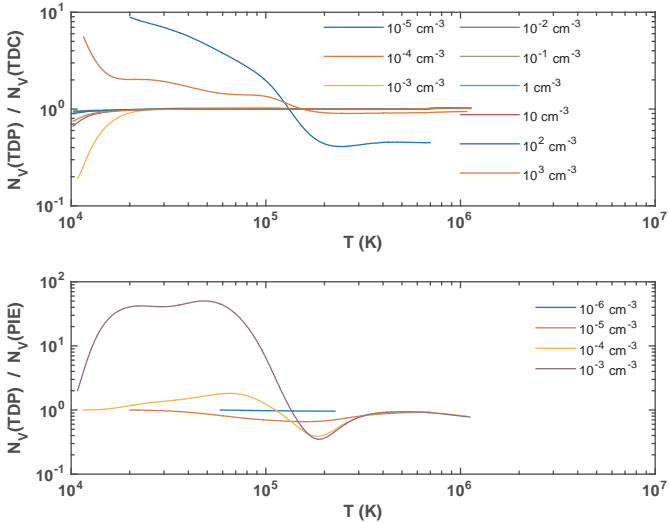}
\caption{Same as Figure~\ref{crat} but for N$^{4+}$.}
\label{nrat}
\end{figure}

\begin{figure}
\epsscale{1.2}
\plotone{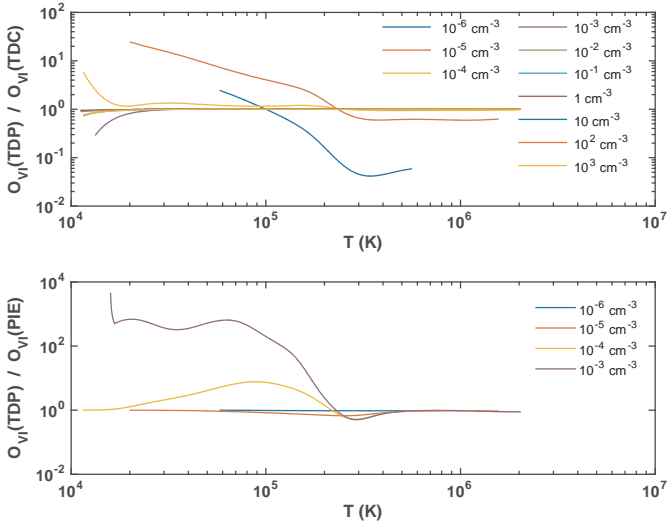}
\caption{Same as Figure~\ref{crat} but for O$^{5+}$.}
\label{orat}
\end{figure}

\begin{figure}
\epsscale{1.2}
\plotone{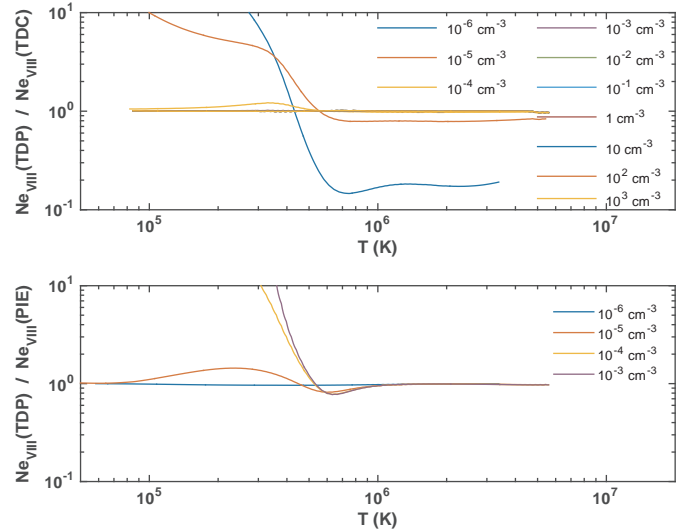}
\caption{Same as Figure~\ref{crat} but for Ne$^{7+}$.}
\label{nerat}
\end{figure}

Figures~\ref{crat}-\ref{nerat} display the ratio between 
the TDP ion fractions and the TDC ion fractions (top panels)
of the dominant coolants  
C$^{3+}$, N$^{4+}$, O$^{5+}$, and Ne$^{7+}$ (for clarity, 
data is only presented 
when the absolute abundance in the TDP gas is 
$>10^{-3}$).
The ratios for different gas densities are displayed by 
different curves.
As expected, the TDP$/$TDC ion fraction of these species
is $\sim$unity for all gas densities $\gtrsim10^{-4}$~cm$^{-3}$, 
expect at $T\lesssim2\times10^4$~K.

The bottom panels of Figures~\ref{crat}-\ref{nerat} display the 
ratio between the TDP ion fractions and the PIE ion fractions for
the same species. These Figures demonstrate that the ion fraction
ratio approaches unity for densities $\lesssim10^{-5}$~cm$^{-3}$.

The convergence of the TDP ion abundance of the dominant coolants 
to the TDC abundances above the threshold density, and to the PIE
abundances below the threshold density, confirm the results of 
section~\ref{cooling}, and verify that the cooling
efficiencies are only affected by both departures from equilibrium
and photoionization at densities very close to the threshold density.

However, as discussed above, the TDP ion fractions of certain species -- 
which do not contribute significantly to the gas cooling -- may deviate
by factors of a few from the TDC or PIE ion fractions.
Therefore, while the thermal evolution of TDP cooling gas may
be considered to be either in PIE or TDC cooling, depending on the 
density, the abundance of specific species must be followed with
a full time-dependent photoionized calculation.

\section{Diagnostics}
\label{sdiagnostics}

Ion-ratios may be used as probes of the ionization state,
temperature, metallicity, and density in photoionized cooling gas.
For a uniform gas cloud,
the column density ratio of the species $m_i$ and $n_j$ may be expressed as 
$\frac{N^m_i}{N^n_j} = \frac{A_m x_i(T)}{A_n x_j(T)}$ where $A_e$ 
is the abundance of element $e$, and $x(T)$ are the ion fractions
computed above.  

\begin{figure}
\epsscale{1.2}
\plotone{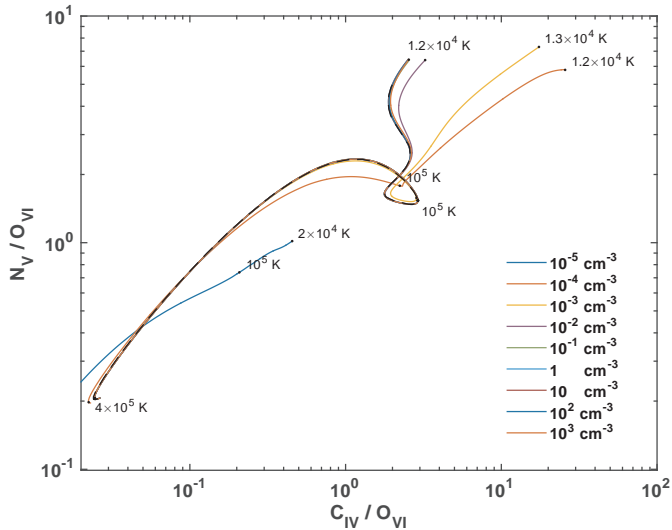}
\caption{Column density ratios 
$N({\rm C}_{\rm IV}) / N({\rm O}_{\rm VI})$
versus $N({\rm N}_{\rm V}) / N({\rm O}_{\rm VI})$, 
for solar metallicity gas exposed to the $z=0$ metagalactic radiation. 
Different solid curves are for different gas densities, as indicated in the 
legend. The dashed curve is for TDC cooling.
As the gas cools, the column density ratios evolve from the 
lower left to the upper right (GS07). Some temperatures along the cooling trajectories
are indicated by the dots and labels along the curves.}
\label{diagnostics}
\end{figure}

As an example, I present in Figure~\ref{diagnostics} ``cooling 
trajectories'' for $N({\rm C}_{\rm IV}) / N({\rm O}_{\rm VI}$)
versus $N({\rm N}_{\rm V}) / N({\rm O}_{\rm VI}$),
for solar metallicity gas exposed to the $z=0$ metagalactic radiation,
for temperatures between $10^6$~K and the thermal equilibrium temperature.
Different gas densities are presented in Figure~\ref{diagnostics}
by different curves (as indicated by the legend). 
For these three ions, O$^{5+}$ is the most highly
ionized, and so $N({\rm C}_{\rm IV}) / N({\rm O}_{\rm VI})$
and $N({\rm N}_{\rm V}) / N({\rm O}_{\rm VI})$ are small at 
high temperature and become large at low temperature. The absorption
line ratios therefore evolve from the lower left, to the upper
right. For reference, some temperature legends are displayed along
the curves (dots, labels). 

The cooling trajectory for TDC cooling gas is displayed by the black 
dashed curve. Figure~\ref{diagnostics} show that for these ratios, the cooling trajectories
for TDP cooling gas with densities $\ge10^{-1}$~cm$^{-3}$ are identical to
the TDC cooling trajectory. At lower densities the photoionizing
radiation can be seen to affect the ion ratios, especially at low temperatures,
as discussed in \S\ref{ionization}. Diagnostic diagrams for other species
of interest may be constructed by using the online data listed in 
Tables~\ref{eqion} and~\ref{tdion} above.

\section{Summary}
\label{summary}

In this paper I present computations of the equilibrium and non-equilibrium
cooling efficiencies, heating rates, and ionization states of low-density radiatively cooling
gas exposed to external photoionizing radiation.
I consider gas containing the 
elements H, He, C, N, O, Ne, Mg, Si, S, and Fe. 
In the calculations I assume cooling in the presence of the 
metagalactic background radiation (HM12) at redshifts $0$ to $3$,
which photoionizes the gas and provides heating.
I present results for gas temperatures, T, between $10^4$ and $10^8$~K,
and for gas densities between $10^{-7}$ and $10^3$~cm$^{-3}$. 
I assume that the gas is dust free, and I consider metallicities Z ranging
from $10^{-3}$ to $2$ times solar. I
carry out computations including rate coefficients for
all relevant atomic recombination and ionization processes, and the
energy loss and gain mechanisms.

For non-equilibrium cooling,
I solve the coupled time-dependent ionization and energy-budget
equations for the cooling gas. For such gas I assume that the
cooling is from an initially hot photoionization equilibrium state
down to its thermal equilibrium temperature. 
The equations and numerical method are presented in \S\ref{method}
and Appendix~\ref{numericaldetails}. 
I calculate the nonequilibrium cooling efficiencies assuming isochoric
evolution (constant density), and discuss the case of isobaric
evolution in \S3.1.

Because I include photoionization by external radiation,
both the equilibrium and time-dependent properties depend
on the assumed density. 
The radiative cooling efficiency at a given temperature is 
therefore a function of both the metallicity $Z$ and density $n_{\rm H}$
(or, equivalently, ionization parameter $U$).
The results of the cooling computations are presented in 
Figures~\ref{fz0}-\ref{fz3}, and listed in Tables~\ref{cool-eq} and
\ref{cool-neq} (see Table~\ref{Tguide} for guidance).

An important conclusion from \S\ref{cooling} is that for every
metallicity there exists
a threshold density (or, equivalently, a threshold ionization
parameter) below which the cooling efficiencies are close to 
those in photoionization equilibrium, and above which they
resemble those in nonequilibrium gas with no external 
radiation.
Above the threshold density, departures from equilibrium ionization
are significant, but the impact of the photoionizing radiation
may be neglected.
Below this threshold density, photoionization plays a critical role,
but departures from equilibrium may be neglected.

The threshold ionization parameter corresponding to $Z\gtrsim1Z_\odot$ is $U=0.42$.
For $z=0$, this corresponds to a threshold density of $10^{-5}$~cm$^{-3}$.
For metallicities $\lesssim10^{-1}~Z_\odot$ the threshold ionization parameter
is a factor of $100$ lower than for solar metallicity ($U=4.2\times10^{-3}$), 
and the threshold densities are therefore a factor of $100$ higher 
($10^{-3}$~cm$^{-3}$).
For higher redshifts, the densities corresponding to these ionization parameter 
are modified as the spectral energy distribution of the metagalactic background 
evolves. The threshold densities corresponding to $U=0.42$ at higher redshifts 
are displayed in Figure~\ref{ncrit}.

The results of the equilibrium and nonequilibrium ionization calculations 
are presented in \S\ref{ionization}. While the thermal evolution (and
the associated abundances of the major coolants) in TDP cooling gas
may be considered to be either in PIE or TDC, depending on the density,
the 
abundance of some specific species may 
differ by factors of a few from the TDC / PIE values,
and must be followed with a full TDP cooling computation.
The full data for 
the time-dependent photoionized ion fractions are listed in online
Tables~\ref{eqion} and \ref{tdion}, as summarized in Table~\ref{ionguide}.

Ion ratios are useful as diagnostic probes. In \S\ref{sdiagnostics} 
I discuss one example, 
$N({\rm C}_{\rm IV}) / N({\rm O}_{\rm VI}$)
versus $N({\rm N}_{\rm V}) / N({\rm O}_{\rm VI}$),
and show how this ratio evolves in photoionized cooling gas and 
how it varies with density for realistic nonequilibrium conditions.

\appendix
\section{Appendix A. Numerical Details}
\label{numericaldetails}

The computations included in this work rely on the numerical scheme outlined in 
GS07 and improved in Gnat \& Sternberg (2009). The details of this
method are summarized here for completeness.

In this work, I consider all ionization stages of the elements 
H, He, C, N, O, Ne, Mg, Si, S, and Fe. 
The microphysical ionization and recombination processes included 
in this computation are
photoionization (Verner et al. 1996), including multielectron Auger 
ionization processes (Kaastra \& Mewe 1993) induced by X-ray photons,
collisional ionization by thermal electrons (Voronov 1997), 
radiative recombination (Aldrovandi \& Pequignot 1973; Shull \& van Steenberg 1982; Landini \& Monsignori Fossi 1990; Landini \& Fossi 1991; Pequignot et al. 1991; Arnaud \& Raymond 1992; Verner et al. 1996), 
dielectronic recombination (Aldrovandi \& Pequignot 1973; Arnaud \& Raymond 1992; Badnell et al. 2003; Badnell 2006; Colgan et al. 2003, 2004, 2005; Zatsarinny et al. 2003, 2004a, 2004b, 2005a, 2005b, 2006; Altun et al. 2004, 2005, 2006; Mitnik \& Badnell 2004), 
and neutralization and ionization by charge transfer reactions with hydrogen and helium atoms and ions (Kingdon \& Ferland fits\footnote{See http://www-cfadc.phy.ornl.gov/astro/ps/data/cx/hydrogen/rates/ct.html.}, based on Kingdon \& Ferland 1996; Ferland et al. 1997; Clarke et al. 1998; Stancil et al. 1998; Arnaud \& Rothenflug 1985), including the statistical charge transfer rate coefficients for high-ions with charges greater than $+4$ (Kingdon \& Ferland 1996). 

The time-dependent equations for the ion abundance
fractions, $x_i$, of element $m$ in ionization stage $i$,
are
\begin{equation}
\label{ion-neq}
\begin{array}{l}
\displaystyle{\frac{dx_i}{dt}} = x_{i-1}~~[q_{i-1}n_{\rm e} + k^{\rm H}_{\uparrow i-1}n_{\rm H^+}
+ k^{\rm He}_{\uparrow i-1}n_{\rm He^+}]
+\displaystyle{\sum_{j<i}}x_j\Gamma_{j\rightarrow i}\\
\;\;\;\;\;\;\;\;\;\;\; + x_{i+1}~~[\alpha_{i+1}n_{\rm e} +
k^{\rm H}_{\downarrow i+1}n_{\rm H^0}
+ k^{\rm He}_{\downarrow i+1}n_{\rm He^0}] \\
\;\;\;\;\;\;\;\;\;\;\; - x_{i}~~[(q_{i} + \alpha_{i})n_{\rm e} + \Gamma_i +
k^{\rm H}_{\downarrow i}n_{\rm H^0}
+ k^{\rm He}_{\downarrow i}n_{\rm He^0}\\
\;\;\;\;\;\;\;\;\;\;\;\;\;\;\;\;\;\;\;\;\; + k^{\rm H}_{\uparrow i}n_{\rm H^+}
+ k^{\rm He}_{\uparrow i}n_{\rm He^+}] \ \ \ .
\end{array}
\end{equation}
In this expression, $q_i$ and $\alpha_i$ are the 
temperature-dependent
rate coefficients
for collisional ionization and recombination (radiative plus
dielectronic), and
$k^{\rm H}_{\downarrow i}$, $k^{\rm He}_{\downarrow i}$,
$k^{\rm H}_{\uparrow i}$, and $k^{\rm He}_{\uparrow i}$
are the rate coefficients for charge transfer reactions
with hydrogen and helium that lead to ionization or neutralization.
The quantities
$n_{\rm H^0}$, $n_{\rm H^+}$, $n_{\rm He^0}$,
$n_{\rm He^+}$, and $n_{\rm e}$ are the particle densities (cm$^{-3}$)
for neutral hydrogen, ionized hydrogen, neutral helium, singly ionized helium, 
and electrons, respectively. $\Gamma_{j\rightarrow i}$ are the 
local
rates of photoionization 
of ions $j$ which result in the ejection of $i-j$ electrons.
$\Gamma_{i}$ are the total photoionization rates of ions $i$ due
to externally incident radiation.
For each element $m$, the ion fractions $x_i = n_{i,m} / (n_HA_m)$ must
at all times satisfy $\sum x_i=1$.

The ionization equations (\ref{ion-neq}) and energy balance 
equation (\ref{energy}) form a set of $103$ coupled ordinary 
differential equations (ODEs). I use a Livermore ODE solver (Hindmarsh 1983)
to integrate these equations. This solver integrates both stiff
and nonstiff ODE systems, and transitions automatically between stiff and 
nonstiff solution methods as necessary (see http://www.netlib.org/odepack/).
Along the solution, the pressure is advanced in small steps,
$\Delta P=\varepsilon P$, where $P$ is the current gas pressure, and 
$\varepsilon\lesssim0.05$. These small pressure steps are associated with small 
time steps $\Delta t$, using equation~\ref{energy}. 
Over the small time step $\Delta t$, the pressure is assumed to evolve 
linearly with time. The cooling efficiencies 
and heating rates at each point are computed using Cloudy (ver. 13.00), 
given the local nonequilibrium ion-fractions. When integrating,
the local errors for hydrogen, helium, and metal ion-fractions were 
set to be smaller than $10^{-9}$, $10^{-8}$, and $10^{-7}$, respectively. 
The accuracy is higher than that used in GS07, because 
heating is sometimes dominated by trace species. This point is further
discussed in Appendix~\ref{xrayapp}.

I verify convergence by repeating the computation at a higher resolution 
(i.e. smaller $\varepsilon$) and confirming that the resulting cooling and heating
rates and ion fractions as functions of time remain unaltered.
I have carried out the integration of the ionization and energy-balance
equations down to the thermal equilibrium temperature, at which point
cooling will cease. However, if the thermal equilibrium temperature is lower 
than $10^4$~K, I stop the computation at $T_{\rm low} = 10^4$~K, because 
molecular chemistry and dust cooling, which are not included in this work, 
may significantly affect the results at lower temperatures.

\section{Appendix B. The Role of X-ray Heating}
\label{xrayapp}

This Appendix considers the role that extreme X-ray radiation plays in
setting the heating rates in dilute plasmas (e.g. Madau \& Efstathiou 1999,
Hambrick et al. 2011).

In order to examine the importance of high-energy photons for the energy
budget in low-density gases, we considered the PIE heating rates under
the influence of the HM12 metagalactic background radiation.
First I consider the heating rates used throughout this paper, 
namely by the HM12 background up to an energy of $8,330$~keV. Then I
considered the HM12 background truncated at an energy of $41$~keV.
The PIE rates were computed using Cloudy, assuming equilibrium
conditions, and taking into account the Klein-Nishina corrections
(G. Ferland, private communication).

\begin{figure*}
\plotone{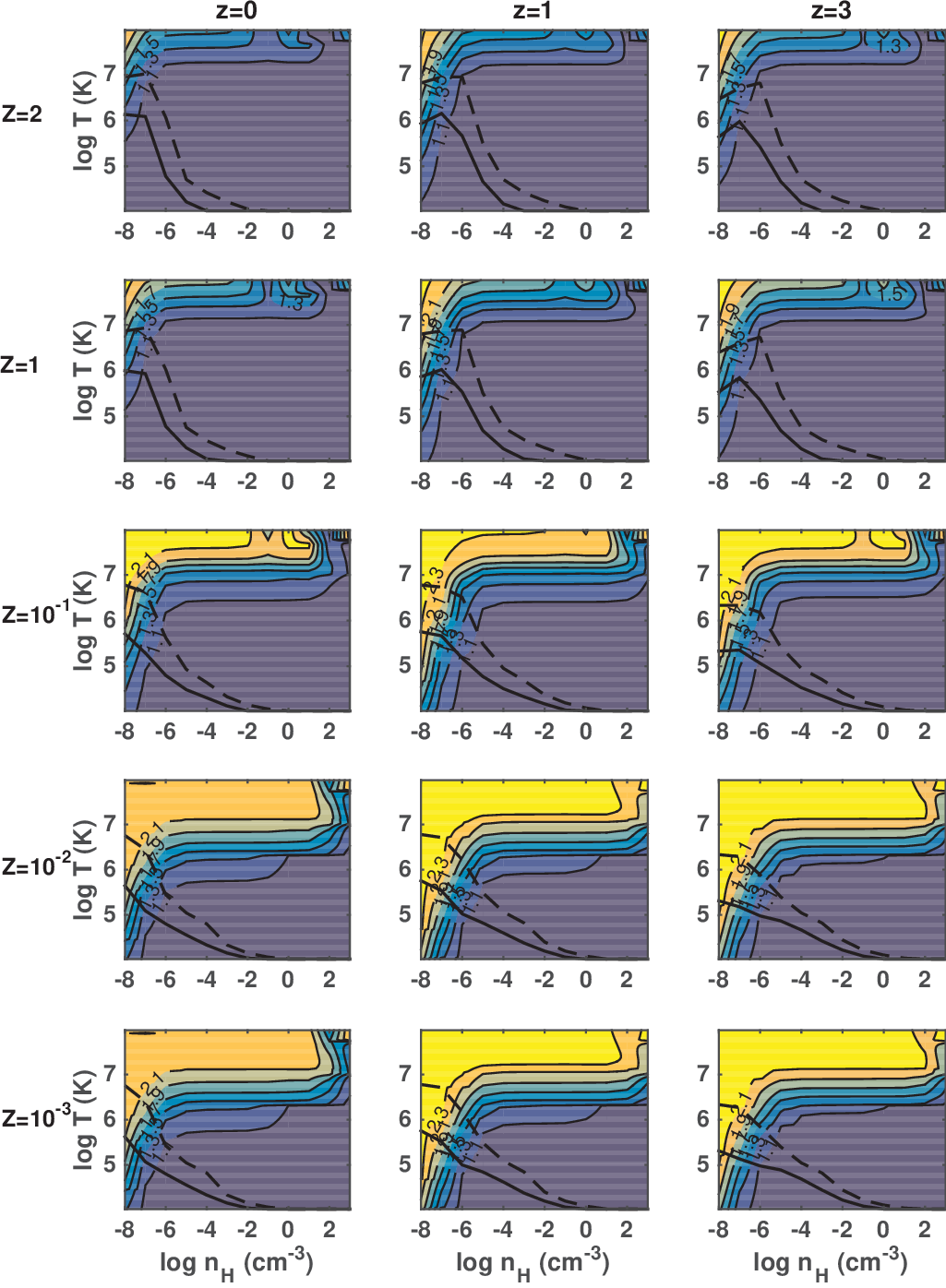}
\caption{The ratio between the heating rate by the $0.12$~eV$ - 8330$~keV HM12
metagalactic background, and the heating rate by the $0.12$~eV$ - 41$~keV HM12
background, for metallicity values $Z=2, 1, 10^{-1}, 10^{-2},$ and $10^{-3}$ times 
solar (from top to bottom), and for the metagalactic background at 
$z=0,1,$ and $2$ (from left to right). Each panel shows the ratio 
$\Upsilon(J_{0.12~{\rm eV} - 8330~{\rm keV}}) / \Upsilon(J_{0.12~{\rm eV} - 41~{\rm keV}})$,
as indicated by the color map, and labeled contours.
The solid black curve in each panel shows the PIE temperatures
(when cooling equals heating) for each density, and the black dashed 
curve shows, for each density, the temperatures at which heating is an
order of magnitude lower than cooling.}
\label{Xrays}
\end{figure*}

In figure~\ref{Xrays}, I plot the ratio between the heating rates,
$\Upsilon(J_{0.12~{\rm eV} - 8330~{\rm keV}}) / \Upsilon(J_{0.12~{\rm eV} - 41~{\rm keV}})$.
I consider five different values of metallicity, $Z=2, 1, 10^{-1}, 10^{-2},$
and $10^{-3}$ time solar, (top to bottom rows), and three different
values of redshift, $z=0, 1,$ and $2$ (from left to right).
For each metallicity-redshift combination, I display the ratio
$\Upsilon(J_{0.12~{\rm eV} - 8330~{\rm keV}}) / \Upsilon(J_{0.12~{\rm eV} - 41~{\rm keV}})$
in PIE
as a function of density (horizontal axis) and temperature (vertical
axis). The ratio is indicated by the color-map, and by the labeled 
contours.

Figure~\ref{Xrays} shows that at low temperatures and high densities
(lower right in each panel)
the X-ray radiation has a negligible contribution to the heating rate.
However, at low densities and high temperatures (upper left), X-ray heating plays
an important role, and including the $40-8,330$~keV radiation, may 
raise the heating rates by factors up to $\sim2$.
For $Z\gtrsim1 Z_\odot$ (two upper rows), extreme-X-ray heating only becomes important
at highly rarified environments (with densities below the cosmic mean,
$n\lesssim10^{-7}$~cm$^{-3}$), and extremely high temperatures 
($T\gtrsim10^7$~K).
For typical WHIM densities and metallicities ($Z\sim0.1Z_\odot$), extreme-X ray heating
contributes to the heating budget for temperatures above a few $\times10^6$~K.
For primordial plasma, it contributes above $\sim10^6$~K.
Figure~\ref{Xrays} shows that in order to accurately estimate the heating
rates in hot dilute plasmas, we need a solid knowledge regarding the intensity
(and SED) of the extreme X-ray radiation.

Note, however, that while the absolute heating rate estimates should
be taken with a grain of salt, our understanding of the thermal evolution
of low-density gas is insensitive to details of the extreme X-ray radiation:
The solid curve in each panel of Figure~\ref{Xrays} shows, for each density,
the photoionization equilibrium temperature at which cooling is exactly 
balanced by heating (computed here using the $0.12 eV - 8330$~keV radiation).
The dashed curves show, for each density, the temperature for which the heating
falls an order of magnitude below cooling. Therefore, in the region to the upper
right of the dashed line, heating is orders of magnitude lower than cooling.
Below the dashed line, the exact value of the heating rate is important 
for the thermal evolution of the gas. In particular, it is important
for determining the value of the equilibrium temperatures.

This demonstrates that within most of the region for which the ratio \newline
$\Upsilon(J_{0.12~{\rm eV} - 8330~{\rm keV}}) / \Upsilon(J_{0.12~{\rm eV} - 41~{\rm keV}})$
is large, the heating has a negligible impact on the thermal evolution of the
plasma because it is orders of magnitude lower than the cooling.
The X-ray radiation therefore only affects the thermal properties of the
plasma when the value of 
$\Upsilon(J_{0.12~{\rm eV} - 8330~{\rm keV}}) / \Upsilon(J_{0.12~{\rm eV} - 41~{\rm keV}})$
is large {\it and} heating is comparable to cooling (i.e. below the dashed
line). Figure~\ref{Xrays} demonstrates that both conditions apply for low
metallicity gas ($Z\lesssim10^{-1} Z_\odot$) at extremely low densities
($n_{\rm H}\lesssim10^{-6}$~cm$^{-3}$). At these densities, the cooling times
are longer than a Hubble time, even for the less energetic SED (truncated
at $40$~keV). For physically relevant parameters (for which the cooling times
are shorter than a Hubble time), the exact details of the X-ray background
are not expected to significantly affect the thermal evolution of the gas.

\section*{Acknowledgments}
This research was supported by the Israeli Centers of Excellence 
(I-CORE) program (center no. 1829/12); by the Israel Science Foundation
(grant No. 857/14); and by program AR-12655 provided by NASA through a 
grant from the STScI, under NASA contract NAS5-26555.


\end{document}